\documentclass[twocolumn]{aastex62}

\newcommand{\sna}{SN~1987A}

\newcommand{\iib}{IIb}
\newcommand{\ma}{M15\nobreakdash-7b}
\newcommand{\mb}{M16\nobreakdash-7b}
\newcommand{\ca}{$^{44}$Ca}
\newcommand{\sa}{$^{44}$Sc}
\newcommand{\ti}{$^{44}$Ti}
\newcommand{\fe}{$^{56}$Fe}
\newcommand{\co}{$^{56}$Co}
\newcommand{\nc}{$^{56}$Ni}
\newcommand{\fr}{$^{57}$Fe}
\newcommand{\cb}{$^{57}$Co}
\newcommand{\nk}{$^{57}$Ni}

\newcommand{\kmps}{~km~s$^{-1}$}

\newcommand{\Msun}{~M$_{\sun}$}

\newcommand{\Zeffsun}{~Z$_{\mathrm{eff,}\sun}$}

\newcommand{\hrich}{H\nobreakdash-rich}
\newcommand{\astrogam}{\mbox{\textit{e-ASTROGAM}}}
\newcommand{\jerkstrand}{A.~Jerkstrand et al.\ 2019, in preparation}
\newcommand{\jerkstrandt}{A.~Jerkstrand et al.\ (2019, in preparation)}
\newcommand{\utrobin}{V.~Utrobin et al.\ 2019, in preparation}
\newcommand{\utrobint}{V.~Utrobin et al.\ (2019, in preparation)}
\newcommand{\gabler}{M.~Gabler et al.\ 2019, in preparation}
\newcommand{\gablert}{M.~Gabler et al.\ (2019, in preparation)}

\usepackage{amsmath}

\received{June 10, 2019}
\revised{July 17, 2019}
\accepted{\today}
\submitjournal{ApJ}
\shorttitle{X-Rays \& Gamma-Rays from Models and SN~1987A}
\shortauthors{Alp et al.}
\begin{document}
\title{X-Ray and Gamma-Ray Emission From Core-collapse Supernovae:
  Comparison of Three-dimensional Neutrino-driven Explosions With
  SN~1987A}

\correspondingauthor{Dennis Alp}
\email{dalp@kth.se}

\author[0000-0002-0427-5592]{Dennis Alp}
\affiliation{Department of Physics, KTH Royal Institute of Technology, 
  The Oskar Klein Centre, AlbaNova, SE\nobreakdash-106\nobreakspace 91 Stockholm, Sweden}

\author[0000-0003-0065-2933]{Josefin Larsson}
\affiliation{Department of Physics, KTH Royal Institute of Technology, 
  The Oskar Klein Centre, AlbaNova, SE\nobreakdash-106\nobreakspace 91 Stockholm, Sweden}

\author[0000-0003-2611-7269]{Keiichi Maeda}
\affiliation{Department of Astronomy, Kyoto University,
  Kitashirakawa-Oiwake-cho, Sakyo-ku, Kyoto, 606-8502, Japan}

\author[0000-0001-8532-3594]{Claes Fransson}
\affiliation{Department of Astronomy, Stockholm University, 
  The Oskar Klein Centre, AlbaNova, SE\nobreakdash-106\nobreakspace 91 Stockholm, Sweden}

\author[0000-0001-8400-8891]{Annop Wongwathanarat}
\affiliation{Max Planck Institute for Astrophysics, 
       Karl-Schwarzschild-Str.~1, D\nobreakdash{-}85748 Garching, Germany}

\author[0000-0002-1663-4513]{Michael Gabler}
\affiliation{Max Planck Institute for Astrophysics, 
       Karl-Schwarzschild-Str.~1, D\nobreakdash{-}85748 Garching, Germany}

\author[0000-0002-0831-3330]{Hans-Thomas Janka}
\affiliation{Max Planck Institute for Astrophysics, 
       Karl-Schwarzschild-Str.~1, D\nobreakdash{-}85748 Garching, Germany}

\author[0000-0001-8005-4030]{Anders Jerkstrand}
\affiliation{Max Planck Institute for Astrophysics, 
       Karl-Schwarzschild-Str.~1, D\nobreakdash{-}85748 Garching,
       Germany}

\author[0000-0002-3684-1325]{Alexander Heger}
\affiliation{Monash Centre for Astrophysics, School of Physics and Astronomy, 
  Monash University, VIC 3800, Australia}
\affiliation{Tsung-Dao Lee Institute, Shanghai 200240, China}

\author[0000-0002-4234-4181]{Athira Menon}
\affiliation{Anton Pannekoek Institute for Astronomy, University of Amsterdam, 1090 GE Amsterdam, The Netherlands}

% 250 words AAS Journals
% 1920 characters arXiv
\begin{abstract}
  During the first few hundred days after the explosion, core-collapse
  supernovae (SNe) emit down-scattered X-rays and gamma-rays
  originating from radioactive line emissions, primarily from the
  \nc{} $\rightarrow$ \co{} $\rightarrow$ \fe{} chain. We use SN
  models based on three-dimensional neutrino-driven explosion
  simulations of single stars and mergers to compute this emission and
  compare the predictions with observations of \sna{}. A number of
  models are clearly excluded, showing that high-energy emission is a
  powerful way of discriminating between models. The best models are
  almost consistent with the observations, but differences that cannot
  be matched by a suitable choice of viewing angle are
  evident. Therefore, our self-consistent models suggest that
  neutrino-driven explosions are able to produce, in principle,
  sufficient mixing, although remaining discrepancies may require
  small changes to the progenitor structures. The soft X-ray cutoff is
  primarily determined by the metallicity of the progenitor envelope.
  The main effect of asymmetries is to vary the flux level by a factor
  of ${\sim}$3. For the more asymmetric models, the shapes of the
  light curves also change. In addition to the models of \sna{}, we
  investigate two models of Type~II\nobreakdash-P SNe and one model of
  a stripped-envelope Type~IIb SN. The Type~II\nobreakdash-P models
  have similar observables as the models of \sna{}, but the
  stripped-envelope SN model is significantly more luminous and
  evolves faster. Finally, we make simple predictions for future
  observations of nearby SNe.
\end{abstract}


%%%%%%%%%%%%%%%%%%%%%%%%%%%%%%%%%%%%%%%%%%%%%%%%%%%%%%%%%%%%%%%%
\section{Introduction}\label{sec:intro}
A core-collapse supernova (CCSN) is the death of a massive star
\citep{baade34, hoyle60}, but the exact nature of the explosion
remains obscured. The so-called delayed neutrino-heating mechanism
\citep{colgate66, arnett66, bethe85, bruenn85} is a leading hypothesis
in which a stalled shock is revived by neutrinos emitted from the
surface of a hot proto-neutron star (for reviews, see
\citealt{janka12, janka17c, burrows13, muller16, janka16,
  couch17}). Recent three-dimensional (3D) simulations are able to
include the basic physics necessary to describe the neutrino
interaction and heating, and to simulate the outcome of the Fe core
collapse, which then connects to the long-term simulations involving
the whole star \citep[][\gabler]{wongwathanarat13,
  wongwathanarat15}. These simulations demonstrated that 3D effects
are important both for the neutrino heating and the hydrodynamic
instabilities above the Fe core.

To verify the supernova (SN) theory and assumptions that go into the
simulations, it is important to compare the model predictions with
observations. The spatial density and abundance distributions of the
ejecta provide key information about the progenitor and explosion
mechanism of \mbox{CCSNe}. Another valuable property that is
observable is the X-ray and gamma-ray emission up to approximately
1000 days after the explosion~(d). This emission arises from the
radioactive decays of the unstable isotopes synthesized during the
first few seconds after core collapse \citep{hoyle54, burbidge57,
  fowler64, hix17}. When an unstable isotope decays, gamma-rays are
emitted. These gamma-rays lose energy due to Compton scattering as
they propagate through the ejecta, and are then either destroyed by
photoelectric absorption or escape the ejecta. An advantage of
studying the early X-ray and gamma-ray emission is that the emission
can be computed directly from the ejecta models without a need of
specifying the microscopic mixing since the thermal conditions are
decoupled from the high-energy radiation field
\citep{jerkstrand11}. The emission is thus a sensitive probe of the
macroscopic mixing and the ejecta structure.  Additionally, the
relevant physics for the photon propagation is well-known and the
gamma-ray transfer is computationally cheap compared to more general
radiation transfer \citep{hillier12, jerkstrand16}.

Several groups have applied this method to compute the X-ray and
gamma-ray emission from CCSN models. Previous studies of 3D models
have explored bipolar Type~II SN models \citep{hungerford03} and
single-lobe Type~II explosion models
\citep{hungerford05}. \citet{maeda06} investigated jet-like
broad-lined SN models and \citet{wollaeger17} computed both optical
and high-energy spectra from a unimodal 3D model. A large number of
models were also created and studied shortly after \object{SN 1987A}
\citep[for an overview, see][]{mccray93}. This early modeling of
\sna{} was based on much more simplified 1D simulations. There are
also several studies that have applied analogous methods to Type~Ia SN
\citep{burrows90, hoflich02, sim07, sim08, kromer09, maeda12, summa13,
  the14} and kilonova models \citep{hotokezaka16, korobkin19}.

In addition to the (down-scattered) X-rays and gamma-rays from the
radioactive decay, other processes can also contribute to the
high-energy emission. Lines from electron transitions are only
relevant below 10~keV in SNe, which is at lower energies than the
component from the radioactive decay. The photons from the radioactive
decay also produce fast recoil electrons through Compton
scattering. The bremsstrahlung emission from these electrons is
expected to be much fainter than the emission from the radioactive
decay in the relevant energy range \citep{clayton91,
  burrows95b}. Interactions with the circumstellar medium (CSM) is
another potential source of X-rays \citep{chevalier94,
  chevalier17}. This is a weak component for the vast majority of all
SNe \citep{dwarkadas14} and is typically at lower X-ray energies,
although the hard X-ray regime is CSM-dominated in some
cases. Examples of such cases include the nearby, strongly-interacting
SN~1993J \citep{leising94, fransson96} and the extremely luminous
Type~IIn SN~2010jl \citep{chevalier12, ofek14}. Such CSM interaction
results in spectra and light curves that are very different from those
produced by reprocessed radioactive decay. Additionally, the gamma-ray
continuum and, in particular, the direct line emission from the
radioactive decay are unlikely to be confused with other emission
components, even under extreme circumstances.

In this paper, we compute the early X-ray and gamma-ray emission from
recent full 3D SN models based on neutrino-driven explosion
simulations \citep{wongwathanarat13, wongwathanarat15} and compare the
predicted emission properties to observations of \sna{}. \sna{} was a
CCSN in the Large Magellanic Cloud (LMC), which makes it the closest
observed SN in more than four centuries (for reviews, see
\citealt{arnett89, mccray93, mccray16}). The proximity of \sna{} makes
it the only CCSN where detailed observations of the early X-ray and
gamma-ray evolution have been possible. The comparisons allow us to
constrain properties of the ejecta structure and composition, and
investigate the viability of recent SN simulations and \sna{}
progenitor models. We also investigate models of other types of
SNe. This allows us to extend the results to more common SN types and
serves as predictions for future observations.

This paper is organized as follows. We describe the SN models in
Section~\ref{sec:mod} and the observations of \sna{} in
Section~\ref{sec:obs}. The algorithm we use for the computations of
the early emission is outlined in Section~\ref{sec:methods} and we
present the results in Section~\ref{sec:results}. We discuss the
results and important details in Section~\ref{sec:discussion} and
provide a summary and the main conclusions in
Section~\ref{sec:conclusions}.

%%%%%%%%%%%%%%%%%%%%%%%%%%%%%%%%%%%%%%%%%%%%%%%%%%%%%%%%%%%%%%%%
\section{Models}\label{sec:mod}
\begin{deluxetable*}{lccccccccc}
\caption{Supernova Explosion Models\label{tab:mod}}
  \tablewidth{0pt}
  \tablehead{\colhead{Model\tablenotemark{a}} & \colhead{Name} & \colhead{Type} & \colhead{$M_\mathrm{ej}$} & \colhead{$t_\text{sim}$\tablenotemark{b}} & \colhead{$E_\mathrm{exp}$} & \colhead{$Z_\mathrm{eff}$\tablenotemark{c}} & \colhead{$\langle v \rangle_{1\,\%}(^{56}\mathrm{Ni})$\tablenotemark{d}} & \colhead{$\langle M_{<} \rangle_{1\,\%}(^{56}\mathrm{Ni})$\tablenotemark{e}}  & \colhead{Ref.} \\
                                   \colhead{} &     \colhead{} &     \colhead{} &    \colhead{(M$_{\sun}$)} &                             \colhead{(d)} &  \colhead{($10^{51}$~erg)} &         \colhead{(Z$_{\mathrm{eff,}\sun}$)} &                                                  \colhead{(km~s$^{-1}$)} &                                                    \colhead{(M$_{\sun}$)} & \colhead{}} \startdata
  B15-1-pw     & B15   & BSG     &            14.2 & 156            & 1.43 & 0.55 & 3530 &            10.5 & 1, 2, 3, 4\\
  N20-4-cw     & N20   & BSG     &            14.3 & 145            & 1.72 & 0.54 & 2110 & \hphantom{1}4.8 & 3, 4, 5, 6, 7, 8\\
  L15-1-cw     & L15   & RSG     &            13.7 & 146            & 1.71 & 0.30 & 4820 &            11.6 & 3, 4, 8, 9\\
  W15-2-cw     & W15   & RSG     &            14.0 & 148            & 1.45 & 0.36 & 4190 &            11.5 & 3, 4, 8, 10\\
  W15-2-cw-IIb & IIb   & He core & \hphantom{1}3.7 & \hphantom{1}18 & 1.52 & 0.36 & 6710 & \hphantom{1}2.8 & 3, 4, 8, 10, 11\\
  M157b-2-pw   & \ma{} & Merger  &            19.5 & \hphantom{11}1 & 1.43 & 0.47 & 3460 &            17.0 & 12, 13\\
  M167b-2-pw   & \mb{} & Merger  &            20.5 & \hphantom{11}1 & 1.41 & 0.47 & 1770 & \hphantom{1}6.7 & 12, 13\\
  \enddata
  \tablerefs{(1) \citet{woosley88}, (2) \citet{bruenn93}, (3)
    \citet{wongwathanarat15}, (4) \gablert{}, (5) \citet{nomoto88}, (6)
    \citet{saio88}, (7) \citet{shigeyama90}, (8)
    \citet{wongwathanarat13}, (9) \citet{limongi00}, (10)
    \citet{woosley95}, (11) \citet{wongwathanarat17}, (12)
    \citet{menon17}, (13) \citet{menon19}}

  \tablenotetext{a}{The first letter does not correspond to any
    physical quantity but is related to the creators. The two-digit
    number is approximately the zero-age main-sequence (ZAMS) mass in
    solar masses. The single-digit number indicates the model number
    in the series of models varying the explosion energy and initial
    seed perturbation \citep{wongwathanarat13}. The last two letters
    are ``pw'' for power-law wind or ``cw'' for constant-wind
    boundary. For the binary merger progenitors, the first two-digit
    numbers give the ZAMS masses of the primary stars in solar masses,
    the following one-digit numbers refer to the ZAMS masses of the
    secondary stars in solar masses, and the last letter before the
    first hyphen is related to the fraction of the He-shell mass
    dredged up.}

  \tablenotetext{b}{The time to which the simulations were
    run. The models are scaled homologously from this time.}

  \tablenotetext{c}{The effective metallicity defined as the
    photoabsorption opacity at 30~keV
    (Section~\ref{sec:progenitor_metallicity}).}

  \tablenotetext{d}{The mass-weighted average radial velocity of the
    fastest 1\,\% of \nc{}.}

  \tablenotetext{e}{The mass-weighted average enclosed mass coordinate
    of the fastest 1\,\% of \nc{}.}
\end{deluxetable*}
We use SN models based on 3D neutrino-driven explosion simulations. An
overview of the models is provided in Table~\ref{tab:mod}. All
progenitor models are 1D but are mapped into three dimensions with
imposed low-amplitude random cell-by-cell perturbations to seed
hydrodynamic instabilities for the SN explosion simulations
\citep{wongwathanarat13}. Below, we describe the properties of the
models that are most relevant to the current study. Comprehensive
descriptions of all simulations can be found in the original
references (Table~\ref{tab:mod}).

All simulations included the neutrino luminosity as a free parameter,
which effectively determines the final explosion energy
($E_\mathrm{exp}$). We note that we do not use the light-bulb
approximation, meaning that the outgoing neutrino luminosities are
considerably modified by the infalling material in our simulations.
The neutrino luminosity was tuned to result in explosion energies of
${\sim}1.5\times{}10^{51}$~erg for all models. This roughly
corresponds to estimates for \sna{} \citep{woosley88, bethe90,
  bethe90b, shigeyama90, blinnikov00, utrobin05}, and is also fairly
representative for ordinary Type~II\nobreakdash-P SNe \citep{kasen09,
  pejcha15, muller17b} and \iib{} SNe \citep{taddia18}.

For a given explosion energy, large variations are expected for the
density and abundance distributions of the ejecta, which are
determined by a coupled interaction between the explosion dynamics and
progenitor structure \citep{wongwathanarat15,utrobin19}. In other
words, these differences can be used to discriminate different
progenitor models using observational data.

The \ma{} and \mb{} models are the results of mergers, whereas all the
other models are single-star models. The B15, N20, \ma{}, and \mb{}
models end their lives as blue supergiants (BSGs) and are designed to
match the progenitor of \sna{}. The B15 model is the single-star
progenitor that yields the best agreement with the optical light curve
of \sna{}, based on self-consistent 3D explosion models
\citep{utrobin15}. For the merger models, \citet{menon19} present
results for artificially mixed 1D explosions while \utrobint{} present
light curve analysis based on the 3D neutrino-driven explosion models.

We have investigated all six merger models of \citet{menon17} that
match the observational properties of the \sna{} progenitor. We only
present results from the \ma{} and \mb{} models here. The \ma{} model
fits the X-ray and gamma-ray observations best and \mb{} is the worst
model. This choice means that the X-ray and gamma-ray properties of
\ma{} and \mb{} roughly bracket those of the complete set of merger
models.

The L15 and W15 models are red supergiants (RSGs), which allow us to
extend the results to Type~II\nobreakdash-P SNe, which is the most
frequent type. Finally, model
W15\nobreakdash-2\nobreakdash-cw\nobreakdash-IIb (\iib{}) is an
explosion of a nearly bare He core with a thin H envelope, which shows
similarity to Cas~A \citep{wongwathanarat17}.

All progenitors except the N20 and \iib{} models were
created using self-consistent stellar evolution simulations. The N20
model \citep{shigeyama90} was created by combining the core
\citep{nomoto88} and envelope \citep{saio88} of two different
models. This was done in an attempt to match the properties of the
progenitor of \sna{}. The \iib{} model was created by artificially
removing all but 0.3\Msun{} of the H envelope of Model
W15\nobreakdash-2\nobreakdash-cw. This model is aimed to mimic a
typical progenitor star of a Type~IIb SN.

The mixing of \nc{} is important for the early X-ray and gamma-ray
emission. In Table~\ref{tab:mod}, we provide the mass-weighted average
radial velocity of the fastest 1\,\% of \nc{}.  This is similar to
\citet{wongwathanarat15}. Here, however, we weight the tracer element
representing the uncertainty in the nucleosynthesis (referred to as X,
Section~\ref{sec:nuc_net}) by 0.5 relative to \nc{}, instead of 1.0,
to remain consistent with the rest of this paper.  The radial velocity
of the fastest \nc{} is a simplified representation of the amount of
mixing in the models. The enclosed mass coordinate of the fastest
\nc{} should be related to the total ejecta mass ($M_\mathrm{ej}$)
because of the importance of the amount of material outside of the
fastest \nc{} (Section~\ref{sec:gen}).

The large range of fastest \nc{} velocities indicates that the set of
explosion models span a large range of mixing properties. The reason
for this has been explored for the B15, N20, L15, and W15 models
\citep{wongwathanarat15, utrobin19}. The mixing in these models is the
result of a complex interplay between the progenitor structure,
dynamics of the SN shock, and propagation of the neutrino-heated
ejecta. One factor that favors efficient mixing is the fast growth of
Rayleigh-Taylor instabilities at the (C+O)/He interface. Furthermore,
a weak interaction of fast Rayleigh–Taylor plumes with the strong
reverse shock occurring below the He/H composition interface also
enhances the amount of mixing. So far, however, studies have relied
solely on a small number of single-star progenitor models. Detailed
analysis of the merger models and comparisons with single-star models
will be presented in a future paper (\utrobin{}).

Finally, we also include some comparisons with the 10HMM model
\citep{pinto88b} as a reference. It was fairly successful at
describing several observables of \sna{} and is representative of the
extensive, but much more simplified, early work on the progenitor of
\sna{} (see Section~2.2 of \citealt{mccray93}). The 10HMM model is 1D
and, importantly, has additional mixing introduced by hand.

\subsection{Geometry}\label{sec:geo}
Of particular relevance to the gamma-ray transfer are the spatial
resolutions of the simulations and the asymmetries caused by the
hydrodynamic instabilities. The late-time simulations were run on
axis-free Yin-Yang grids \citep{kageyama04, wongwathanarat10b} with a
relative radial resolution better than 1\,\% at all radii and an
angular resolution of 2\degr{}. For this study, the models are mapped
from the Yin-Yang grids to spherical grids. The resolution of the
spherical grids is sufficient to perform the mapping without
significant loss of characteristic structures. The simulations provide
no information on the small-scale mixing below the grid scale, but
this does not affect the properties of the ejecta that are relevant
for the gamma-ray propagation.

The \hrich{} single-star models are evolved until ${\sim}$150~d and
the \iib{} model until 18~d, beyond which they can be assumed to
expand homologously (\gabler{}). The merger models are only available
at an age of 1~d, from which we scale the models homologously. We
check the effects of the late-time radioactive heating on the dynamics
by comparing results from a version of the B15 model that has been
expanded homologously from 1~d to the standard B15 model that has been
followed by simulations until 156~d. The primary difference introduced
by the late-time heating is increased mixing, leading to a flux
increase of ${\sim}20$\,\% for most cases, although it can reach
${\sim}$40\,\% for the direct line emission at early times ($<200$~d).
Homologous expansion is a reasonable approximation for a few thousand
years \citep[e.g.,][]{truelove99}, but this period of the evolution
could be more than an order of magnitude shorter in extreme cases,
such as Type~IIn SNe and in the presence of the equatorial ring in the
case of \sna{}. In general, the ejecta expand homologously inside of
the reverse shock resulting from the interaction of the outermost
ejecta with the CSM. During the first 1000~d of \sna{}, before
interaction with the equatorial ring, all of the ejecta can safely be
assumed to expand homologously. In fact, the central parts of the
ejecta that contain the majority of the mass are still freely
expanding at current epochs \citep{fransson13}.

The explosion in all models sets in strongly asymmetrically as a
consequence of hydrodynamic instabilities associated with the
deposition of neutrino energy behind the stalled shock. This happens
during the first seconds of post-bounce evolution. These initial
explosion asymmetries trigger the growth of secondary non-radial
hydrodynamic instabilities after the shock crosses the
composition-shell interfaces on its way out from the center to the
surface of the exploding star. A more detailed description of the
ejecta asymmetries can be found in Section~5 of
\citet{wongwathanarat15} and Section~3.3 of \citet{utrobin19}.

\subsection{Radioactive Elements}\label{sec:nuc_net}
The nucleosynthesis is treated slightly differently in the different
models. The nucleosynthesis in the \sna{} models was followed by a
network of elements that includes $^{1}$H; the 13
$\alpha$\nobreakdash{-}nuclei from $^{4}$He to $^{56}$Ni; and a tracer
nucleus X \citep{kifonidis03, wongwathanarat13, wongwathanarat15}. The
tracer represents elements that were produced in grid cells where the
electron fraction was below 0.49. Therefore, X comprises neutron-rich,
Fe-group elements, but the simulations provide no additional
information about the compositions. The networks used for the RSG and
\iib{} models omit $^{32}$S, $^{36}$Ar, $^{48}$Cr, and $^{52}$Fe.

\begin{deluxetable*}{lcccccccc}
\tablecaption{Nuclear Decay Data\label{tab:ato}}
\tablewidth{0pt}
\tablehead{
    \colhead{Isotope} & \colhead{$\tau$\tablenotemark{a}} &
    \colhead{Ref.\tablenotemark{b}} & \colhead{Lines\tablenotemark{c}}
    & \colhead{Intensity} & \colhead{$E_1$\tablenotemark{d}} & \colhead{$I_1$\tablenotemark{d}} & \colhead{$E_2$\tablenotemark{d}} & \colhead{$I_2$\tablenotemark{d}} \\
    \colhead{} & \colhead{(day)} &
    \colhead{} &
    \colhead{} & \colhead{($\gamma$ decay$^{-1}$)} &
    \colhead{(keV)} & \colhead{($\gamma$ decay$^{-1}$)} &
    \colhead{(keV)} & \colhead{($\gamma$ decay$^{-1}$)}} \startdata
  \nc{} & $\hphantom{00}6.08$   & 1, 2, 3, 4 & \hphantom{0}6\hphantom{\tablenotemark{e}} & $3.21$ & 158 & 0.99 & \hphantom{0}812 & 0.86 \\
  \co{} &  $\hphantom{0}77.24$  & 1, 2, 3, 4 & 45\tablenotemark{e}                       & $2.91$ & 847 & 1.00 &            1238 & 0.68 \\
  \cb{} &              $271.74$ & 1, 5, 6    & 10\hphantom{\tablenotemark{e}}            & $1.06$ & 122 & 0.86 & \hphantom{0}136 & 0.11 \\
 \enddata
  \tablerefs{(1) \citet{firestone99}, (2) \citet{junde92}, (3)
    \citet{junde99}, (4) \citet{junde11}, (5) \citet{bhat92}, (6)
    \citet{bhat98}
  }

  \tablenotetext{a}{Half-life. Divide by $\ln(2)$ for lifetime.}

  \tablenotetext{b}{Reference for the lifetime.}
  
  \tablenotetext{c}{Number of lines. From the Table of Isotopes
    \citep{firestone99}.}

  \tablenotetext{d}{Line energies ($E_i$) and intensities ($I_i$) of
    the strongest ($i=1$) and second strongest ($i=2$) lines. From the
    Table of Isotopes \citep{firestone99}.}
  
  \tablenotetext{e}{Including the positron-annihilation line at
    511~keV.}
\end{deluxetable*}
The most important nuclear decay data are provided in
Table~\ref{tab:ato}.  We follow the decays \nc{} $\rightarrow$ \co{}
$\rightarrow$ \fe{} \citep{clayton69} and \cb{} $\rightarrow$ \fr{}
\citep{clayton74}. The most important transition during the relevant
time intervals in the current context is \co{} $\rightarrow$
\fe{}. The transition \cb{} $\rightarrow$ \fr{} dominates below
100~keV at epochs later than ${\sim}$800~d for the \hrich{} models and
${\sim}$300~d for the stripped IIb model. The contribution from the
\ti{} $\rightarrow$ \sa{} $\rightarrow$ \ca{} chain is negligible
during the time periods considered in this paper. Finally, the isotope
\nk{} has a half-life of 36~h and is much less abundant than
\nc{}. The contribution of \nk{} $\rightarrow$ \cb{} is, therefore,
always negligible.

The true composition of X and the limited nuclear networks introduce
uncertainties in the masses of individual isotopes. We compute the
opacities based on the masses from the simulations, but rescale the
escaping fluxes as if the masses of the radioactive sources match
observations of \sna{} for all models. All \nc{} masses from the
simulations are consistent, within the large uncertainties, with the
${\sim}$0.07\Msun{} of \nc{} produced by \sna{} \citep{suntzeff90,
  bouchet91}. The spatial distribution of \nc{} is taken to be the
distribution of the sum of the \nc{} explicitly followed by the grid
weighted by 1 and the tracer X weighted by 0.5. We investigate the
effect of this choice by comparing results based on weighing the
tracer by 0 and 1. The flux level varies by less than $25$\,\% in most
cases, the shapes of the spectra are practically unchanged, and light
curve peaks shift by less than ${\sim}$50~d.

We set the mass of \nk{} by adopting a fixed ratio of the mass of
\nk{} relative to that of \nc{}.
The estimates of the \nk{}$/$\nc{} ratio of \sna{} from explosive
nucleosynthesis networks \citep{thielemann90, woosley91b}, direct
observations \citep{syunyaev90, kurfess92}, and light curve modeling
\citep{fransson93} favor values around twice the solar ratio, where
the solar system $^{57}$Fe$/\text{\fe}$ number ratio is 0.023
\citep{lodders03}.  The spatial distribution of \nk{} is represented
by the tracer X.

For consistency and for facilitating comparisons, we scale the fluxes
of all models to match the radioactive isotope yields of \sna{}. This
puts the \nc{} masses of our RSG explosions around the 80th percentile
of the distribution of inferred yields of ordinary
Type~II\nobreakdash-P SNe \citep{kasen09, pejcha15, muller17b,
  anderson19}. In contrast, the \nc{} mass of our \iib{} model is at
the 20th percentile of the inferred \nc{} masses of Type~IIb SNe
\citep{taddia18, anderson19}, which peaks at roughly twice the \nc{}
mass of \sna{}.

\subsection{Progenitor Metallicity}\label{sec:progenitor_metallicity}
The progenitor metallicity is important because it determines the
composition of the outermost parts of the models that are not mixed
with the freshly synthesized material. The envelope metallicity
significantly affects the emerging emission because the metals
(primarily Fe) dominate the photoabsorption opacity in the relevant
10--100~keV range, even at metallicities that are much lower than the
solar metallicity.

Instead of using the standard definition of metallicity, we define the
effective metallicity ($Z_\mathrm{eff}$) as the photoabsorption
opacity at 30~keV. This is done because the abundance patterns of the
different models are different and the photoabsorption opacity is the
most important consequence of the metallicity in the current
context. The effective metallicities of all models are provided in
Table~\ref{tab:mod}. It is worth pointing out that the effective
metallicity of the LMC is 0.54\Zeffsun{}. Relative to solar
metallicity, this is slightly higher than what is typically adopted as
the (standard) metallicity of the LMC. This is a result of Fe
dominating the photoabsorption opacity above 6.4~keV and that Fe is
not as under-abundant as many of the intermediate-mass elements.

\begin{deluxetable}{lcccccccccccccc}
\caption{Adopted LMC Abundances\label{tab:abu}}
  \tablewidth{0pt}
  \tablehead{\colhead{Element} & \colhead{LMC\tablenotemark{a}} & \colhead{Solar\tablenotemark{a,b}} & \colhead{Difference} & \colhead{LMC Ref.} \\
             \colhead{}        & \colhead{(dex)}                    & \colhead{(dex)}                & \colhead{(dex)}      & \colhead{}}\startdata
  H  &     $\equiv{}12$ &     $\equiv{}12$ &        \nodata{} & \nodata{} \\
  He &            10.93 &            10.90 & \hphantom{$-$}0.03 & 1, 2 \\
  C  & \hphantom{1}7.81 & \hphantom{1}8.39 &            $-$0.58 & 1, 2, 3, 4 \\
  O  & \hphantom{1}8.35 & \hphantom{1}8.69 &            $-$0.34 & 1, 2, 4, 5 \\
  Ne & \hphantom{1}7.58 & \hphantom{1}7.87 &            $-$0.29 & 1, 2 \\
  Mg & \hphantom{1}7.06 & \hphantom{1}7.55 &            $-$0.49 & 4, 5 \\
  Si & \hphantom{1}7.20 & \hphantom{1}7.54 &            $-$0.34 & 4, 5 \\
  S  & \hphantom{1}6.78 & \hphantom{1}7.19 &            $-$0.41 & 1 \\
  Ar & \hphantom{1}6.48 & \hphantom{1}6.55 &            $-$0.07 & 1 \\
  Ca & \hphantom{1}6.02 & \hphantom{1}6.34 &            $-$0.32 & 1, 3 \\
  Ti & \hphantom{1}4.81 & \hphantom{1}4.92 &            $-$0.11 & 6 \\
  Cr & \hphantom{1}5.42 & \hphantom{1}5.65 &            $-$0.23 & 1, 3 \\
  Fe & \hphantom{1}7.23 & \hphantom{1}7.47 &            $-$0.24 & 1, 3, 5 \\
  \enddata
  \tablerefs{(1) Table~12 of \citet{russell90}, (2) Table~5 of
    \citet{kurt98}; (3) \citet{russell89}; (4) Table~17 of \citet{hunter07b};
    (5) Table~9 of \citet{trundle07}; (6) Table~1 of
    \citet{russell92}}

  \tablecomments{The effective metallicity (see
    Section~\ref{sec:progenitor_metallicity}) of the LMC is
    $0.54$\Zeffsun{}. This is dominated by the difference in the Fe
    abundance of $-0.24$~dex. The solar abundances are included for
    reference.}

  \tablenotetext{a}{Number abundance of element El is represented by
    the astronomical log scale
    $12 + \log_{10}[A(\mathrm{El})/A(\mathrm{H})]$.}

  \tablenotetext{b}{Solar abundances from Table~1 of \citet{lodders03}.}
\end{deluxetable}
The treatment of the metallicity in the stellar evolution simulations
of the B15 and N20 progenitors has been significantly simplified.
Those nuclear networks were reduced by omitting the heavier metals and
representing the metallicity using only lighter elements. We correct
for this by raising the mass fraction of each individual element to
the LMC abundance in grid cells where the individual abundance is
lower than the corresponding LMC value. Abundances are never lowered
to match LMC abundances, which explains why the B15 model
(0.55\Zeffsun{}) has marginally higher effective metallicity than the
LMC (0.54\Zeffsun{}). We do not use abundances inferred from the
equatorial ring of \sna{} because this would require much larger
changes to the models. See Section~\ref{sec:dis_pro_met} for a
discussion of using abundances inferred from the equatorial ring of
\sna{} instead of LMC abundances.

The adopted LMC abundances are provided in Table~\ref{tab:abu}. The
corrections are performed in regions where the H mass fraction is
$>0.1$. The H mass fractions are slightly reduced to preserve the
total masses. The total changes are approximately shifts of 0.1\Msun{}
from H to metals, primarily intermediate-mass elements. We reiterate
that only the B15 and N20 models require this modification. For
example, the effective metallicity of the B15 model before correction
is 0.03\Zeffsun{}. We also note that the RSGs, mergers, and \iib{}
models are unmodified from their evolutionary compositions and have
slightly lower effective metallicities (Table~\ref{tab:mod}).

%%%%%%%%%%%%%%%%%%%%%%%%%%%%%%%%%%%%%%%%%%%%%%%%%%%%%%%%%%%%%%%%
\section{Observations of \sna{}}\label{sec:obs}
We use early X-ray and gamma-ray observations of \sna{} as an
observational test of the simulated SN models. The comparisons can be
divided into three categories; spectra, continuum light curves, and
line fluxes. The observations of the line profiles are investigated in
a separate paper (\jerkstrand{}).

The distance to \sna{} is taken to be 51.2~kpc \citep{panagia91,
  gould98, panagia99, mitchell02} and when comparing simulations to
observations, we correct the observed spectra for the recessional
heliocentric velocity of the LMC of 287\kmps{} \citep{groningsson08b,
  groningsson08}. The ISM absorption is negligible in the relevant
energy range of ${\sim}$10--3500~keV (e.g., \citealt{willingale13,
  frank16}).

\begin{deluxetable*}{lccccccccc}  
  \tablecaption{Early Hard X-Ray and Gamma-Ray Observations of \sna{}\label{tab:obs}}
  \tablewidth{0pt}
  \tablehead{\colhead{Instrument\tablenotemark{a}} &
    \colhead{Platform} & \colhead{Epoch} &
    \colhead{Energy Range} & \colhead{References} \\
    \colhead{} & \colhead{} & \colhead{(d)} & \colhead{(keV)} & \colhead{}}\startdata
  HEXE                       & Roentgen/\textit{Mir-Kvant} & 168--830          &  15--200    & 1, 2, 3, 4, 5, 6 \\
  Pulsar X-1                 & Roentgen/\textit{Mir-Kvant} & 168--830          &  70--600    & 1, 2, 3, 4, 5, 6 \\
  GRS                        & \textit{SMM}                & 1--826            & 300--9000   & 7, 8, 9, 10 \\
  LAC                        & \textit{Ginga}              & 2--1400           &   6--28     & 11, 12, 13, 14, 15 \\
  MSFC/Lockheed/Marshall     & Balloon                     & 95, 249, 411, 619 &  18--960    & 16, 17, 18, 19, 20 \\
  GRIP/Caltech/CIT           & Balloon                     & 86, 268, 414, 771 &  30--5000   & 21, 22, 23 \\
  GRIS/GSFC/Bell/SNLA/Sandia & Balloon                     & 433, 613          &  20--8000   & 24, 25, 26, 27, 28 \\
  JPL                        & Balloon                     & 286               &  50--8100   & 29 \\
  GRAD/Florida/GSFC          & Balloon                     & 319               & 700--3000   & 30 \\
  UCR Compton Telescope      & Balloon                     & 418               & 500--20,000 & 31, 32, 33 \\
  \enddata
  \tablerefs{(1) \citet{sunyaev87}, (2) \citet{syunyaev87}, (3) \citet{syunyaev88}, (4) \citet{syunyaev89}, (5) \citet{syunyaev90}, (6) \citet{sunyaev91}, (7) \citet{forrest80}, (8) \citet{matz88}, (9) \citet{leising89}, (10) \citet{leising90}, (11) \citet{makino87}, (12) \citet{dotani87}, (13) \citet{tanaka88}, (14) \citet{tanaka91}, (15) \citet{inoue91}, (16) \citet{sandie88}, (17) \citet{sandie88b}, (18) \citet{wilson88}, (19) \citet{fishman90}, (20) \citet{pendleton95}, (21) \citet{althouse85}, (22) \citet{cook88}, (23) \citet{palmer93}, (24) \citet{teegarden89}, (25) \citet{tueller90}, (26) \citet{tueller91}, (27) \citet{tueller91b}, (28) \citet{barthelmy91}, (29) \citet{mahoney88}, (30) \citet{rester89}, (31) \citet{zych83}, (32) \citet{simone85}, (33) \citet{ait_ouamer92}}
  \tablenotetext{a}{The abbreviations and acronyms are:
    High Energy X-ray Experiment (HEXE),
    Gamma-Ray Spectrometer (GRS),
    \textit{Solar Maximum Mission} (\textit{SMM}),
    Large Area Proportional Counter (LAC),
    Marshall Space Flight Center (MSFC),
    Gamma-Ray Imaging Payload (GRIP),
    California Institute of Technology (CIT),
    Gamma-Ray Imaging Spectrometer (GRIS),
    Goddard Space Flight Center (GSFC),
    Sandia National Laboratories Albuquerque (SNLA),
    Jet Propulsion Laboratory (JPL),
    Gamma-Ray Advanced Detector (GRAD),
    University of California, Riverside (UCR).
    }
\end{deluxetable*}
An overview of all early hard X-ray and gamma-ray observations of
\sna{} is provided in Table~\ref{tab:obs}. Some of the early X-ray and
gamma-ray observations have also been summarized by other authors
\citep{bunner88, gehrels88, leising91, teegarden91, tueller91b,
  wamsteker93}. Below, we briefly describe the instruments and data
used for our comparisons.

\subsection{The Roentgen Observatory}
The Roentgen Observatory was an experiment in the \textit{Kvant}
module of the space station \textit{Mir}. The \textit{Kvant} module
carrying Roentgen docked to \textit{Mir} during 1987 April and started
observing \sna{} 168~d after the outburst and continued to monitor
\sna{} to later than 800~d. We use data from the High Energy X-ray
Experiment (HEXE) and the Pulsar~X\nobreakdash{-}1\footnote{Not to be
  confused with the bright X-ray source LMC~X\nobreakdash{-}1, which
  is located 0.6\degr{} from \sna{}.} instruments.

HEXE was the most sensitive instrument and provides relatively
accurate continuum light curves in three energy bands in 15--200~keV
\citep{syunyaev90}. Low-resolution spectra were also extracted for
seven epochs. We also use spectra from Pulsar~X\nobreakdash{-}1 at
320~d \citep{syunyaev88}. The data are of significantly lower quality
and are primarily included with the purpose of verifying the HEXE
spectra above 70~keV during peak brightness, while also extending the
energy coverage to 600~keV.

\subsection{\textit{SMM}}
The \textit{Solar Maximum Mission} (\textit{SMM}) was launched in 1980
and was a dedicated solar observatory. One of the seven instruments on
board was the Gamma-Ray Spectrometer (GRS), which operates in the
energy range 0.3--8.5~MeV \citep{forrest80}. The moderate energy
resolution of around 50~keV at 1~MeV is insufficient for resolving the
radioactive line profiles. \textit{SMM} was unable to point away from
the Sun except for short durations, so \sna{} was observed as a part
of the background emission while \textit{SMM} performed solar
observations. Because \textit{SMM} was operational for seven years
before \sna{}, it was possible to monitor the background gamma-ray
flux before the explosion and then detect \sna{} as a change in the
background level during its early evolution.

We use data from \textit{SMM} to constrain the evolution of the line
fluxes. \textit{SMM} provides the most accurate measurements of the
line fluxes and also continuously monitored \sna{} throughout its
early evolution. No emission from \sna{} was detected before 1987
July, but the \co{} decay lines rose rapidly during 150 to 200~d
\citep{matz88}. The line emission peaked shortly thereafter, and then
decayed beyond the detection threshold around day 600
\citep{leising90}.

\subsection{\textit{Ginga}}
The X-ray astronomy satellite \textit{Ginga} (``Galaxy'') was launched
on 1987 February 5 and started observing \sna{} two days after
outburst and monitored \sna{} for more than 1000 days. The first
detection of emission from \sna{} was on day 131 \citep{dotani87}.
\textit{Ginga} carried three instruments, but only data from the Large
Area Proportional Counter (LAC) are used for this work. The LAC was a
collimator and provided data on \sna{} in the 6--28~keV range. Because
of the relatively large collimator opening angle, the data might be
contaminated by background sources \citep{inoue91}. For this reason,
we choose to exclude data during an apparent flaring period in January
1988.

We use the \textit{Ginga} light curves in the 6--16 and 16--28~keV
ranges. For the presentations in this paper, we treat the two bands as
spectral bins for the comparisons at specific times. The X-ray
emission in the full 6--28~keV range consists of two emission
components \citep{tanaka88} and the energy cut at 16~keV was chosen to
optimally separate them from each other \citep{inoue91}. The
signal-to-noise ratio in the low-energy band is low and the emission
could be the result of interactions with the CSM or bremsstrahlung
from the fast recoil electrons. This means that the low-energy band is
not strictly comparable to our predictions but could be viewed as an
upper limit.  The 16--28~keV component is less variable and most
likely represents the low-energy cutoff of the Comptonized radioactive
emission \citep{inoue91, tanaka91}. The \textit{Ginga} data agree
reasonably well with the Roentgen/HEXE data in the overlapping energy
range.

\subsection{Balloon-borne Experiments}
We use data from a number of balloon-borne experiments, the details of
which are summarized in Table~\ref{tab:obs}. They provide independent
measurements of the spectra (MSFC, GRIP), which agree well with the
data from HEXE and Pulsar~X\nobreakdash{-}1. This effectively also
verifies the continuum flux measurements of HEXE that are used to
construct the continuum light curves. All of the balloon experiments
that we include were able to measure line fluxes, which show
reasonable agreement with the line flux evolution as observed by
\textit{SMM} (Figure~5 of \citealt{leising90}). Additionally, only
balloon-borne experiments (GRIS, JPL, GRAD) carried instruments with
sufficient energy resolution to resolve the line profiles of \co{}
(presented in \jerkstrand{}).

%%%%%%%%%%%%%%%%%%%%%%%%%%%%%%%%%%%%%%%%%%%%%%%%%%%%%%%%%%%%%%%%
\section{Algorithm \& Implementation}\label{sec:methods}
We compute the X-ray and gamma-ray properties by following the
propagation of Monte Carlo (MC) photons through the ejecta. The
interactions we include are photoelectric absorption, Compton
scattering, and pair production. We do not compute bremsstrahlung or
fluorescent emission because these effects only contribute
significantly at energies below the sharp photoabsorption cutoff
around 15~keV \citep{woosley89, clayton91, burrows95b}.
We verify our code by comparing to previously published spectra and
light curves \citep{milne04, the14} of the W7 model \citep{nomoto84b},
and by comparing results from our code to those from \texttt{HEIMDALL}
\citep{maeda14} and \texttt{SUMO\nobreakdash-3D} (\jerkstrand{}).

We use $\xi_i$ to denote random numbers uniformly distributed in the
range 0 to 1, where the index $i$ distinguishes the different random
numbers. Spatial vectors are denoted using boldface (e.g.,
$\boldsymbol{a}$), and the Cartesian components are $(a_x, a_y, a_z)$.
The spherical components are $(a_\rho{}, a_\phi{}, a_\theta{})$, where
$\phi{}$ is the azimuthal angle ranging from 0 to $2\,\pi$, and the
polar angle $\theta$ is defined from 0 at the north pole to $\pi$ at
the south pole. Four-vectors are denoted using Greek indices (e.g.,
$b_\mu$). We denote quantities in the SN frame without primes and
quantities in frames locally comoving with the homologous expansion
with primes.

\subsection{Cross Sections}
In this section, we describe the cross sections used for Compton
scattering, photoabsorption, and pair production. The absorption and
scattering properties of the tracer X are set to those of $^{56}$Fe.

\subsubsection{Compton Scattering}
The Compton scattering cross section is given by the Klein-Nishina
formula (e.g., \citealt{rybicki79}). For computational stability and
performance, we use an approximate expression for the Klein-Nishina
cross section \citep[p.\ 319 of][]{pozdnyakov83}. The difference
between the approximation and the exact formula is less than 1\,\% at
all energies.

The energy lost by a photon in each scattering in the rest frame of
the electron is given by
\begin{equation}
  \label{eq:frac_ene}
  \frac{E_\mathrm{f}^\prime}{E_\mathrm{i}^\prime}=\frac{1}{1+\frac{E_\mathrm{i}^\prime}{m_\mathrm{e}\,c^2}(1-\cos\alpha^\prime)},
\end{equation}
where $E_\mathrm{f}$ is the final photon energy, $E_\mathrm{i}$ the
initial photon energy, $m_\mathrm{e}$ the electron mass, and $\alpha$
the scattering angle. It shows the important property that the
fractional energy loss is much higher when the energy before
scattering is comparable to, or higher than, the electron rest
energy.

\subsubsection{Photoelectric Absorption}
Photoabsorption cross sections are quickly declining functions of
energy $E$, which implies that photoabsorption is only relevant at
relatively low X-ray energies. It starts becoming important around
100~keV and dominates below ${\sim}$40~keV for typical SN abundances
\citep{alp18c}. The non-relativistic high-energy asymptote is
approximately proportional to $E^{-3}$. We use the cross sections of
\citet{verner96}.

We note that the cross sections of \citet{verner96} are very similar
to those of \citet{verner95} at energies above 10~keV for the relevant
isotopes, except for $^4$He. The photoabsorption cross section of
$^4$He from \citet{verner96} is approximately 40\,\% lower than the
value of \citet{verner95} but the effect on the escaping fluxes is
only approximately 1\,\%. The cross sections of \citet{verner96} are
also very similar to those of \citet{veigele73}.\footnote{In passing,
  we also note that the photoelectric cross sections of H and He have
  been switched in Table~3 of \citet{hoeflich92}.}

\subsubsection{Pair Production}
We include pair production primarily for comparisons with other codes
\citep{milne04}. We have indeed verified that this effect has a
negligible impact on the results for the models studied in this paper.
For example, for \nc{}, the scattering cross section is 10 times
higher than the pair production cross section at 3~MeV and they are
approximately equal at 10~MeV \citep[see Figure~1
of][]{milne04}. Typically, less than 0.1\,\% of the total photons are
pair absorbed.

We take the pair production cross sections from \citet{ambwani88},
which interpolates tabulated values of the cross sections
\citep{hubbell69, hubbell80}. The pair production cross sections are
\begin{widetext}
  \begin{equation}
    \label{eq:ppxs}
    \sigma_\text{pp} = \begin{cases}
      0\text{~cm}^2 & E_6 < 1.022\text{~MeV} \\
      0.10063\,(E_6-1.022)\,Z^2\times 10^{-27}\text{~cm}^2 & 1.022 \leq E_6 < 1.5\text{~MeV} \\
      [0.0481+0.301\,(E_6-1.5)]\,Z^2 \times 10^{-27}\text{~cm}^2 & E_6 \geq 1.5\text{~MeV}
    \end{cases},
  \end{equation}
\end{widetext}
where $E_6$ is the photon energy in units of MeV and $Z$ is the atomic
number.\footnote{As noted by \citet{swartz95}, the factor is supposed
  to be 0.10063, not 1.0063 as printed in Eq.~(2) of
  \citet{ambwani88}.}

\subsection{Photon Initialization}
Gamma-ray photons are created by the decay of the radioactive isotopes
in the inner regions of the ejecta. Each MC photon is initially
created by assigning a spatial position $\boldsymbol r$, a direction
of propagation $\boldsymbol{\Omega}$, and a photon energy. The initial
spatial distribution is given by the spatial distribution of the
parent nuclei in the models (Section~\ref{sec:nuc_net}). The direction
of the photons are random in the rest frame of the homologously
expanding parent nuclei, and the initial photon energies are the
well-defined energies of the corresponding transitions. Important data
about the radioactive lines included in the code are provided in
Table~\ref{tab:ato}. With the energy $E^\prime$ and the direction
$\boldsymbol{\Omega^\prime}$, it is straightforward to construct the
four-momentum $p_\mu^\prime$ and inverse Lorentz boost it to the SN
frame. The result of accounting for the difference in reference frames
is that the energies $E$ are Doppler shifted and $\boldsymbol{\Omega}$
slightly beamed in the outward direction. The effect on the result
introduced by the beaming is an increase in escaping number flux of
less than 1\,\%.

\subsection{Photon Propagation}
As the photon propagates through the ejecta, the optical depth
is given by
\begin{equation}
  \label{eq:int_tau}
  \tau = \tau_\mathrm{CS}+\tau_\mathrm{ph}+\tau_\mathrm{pp},
\end{equation}
where the scattering depth $\tau_\mathrm{CS}$, photoabsorption depth
$\tau_\mathrm{ph}$, and pair production depth $\tau_\mathrm{pp}$ are
given by
\begin{align}
  \label{eq:tau_cs}
  \tau_\mathrm{CS}&=
  \int_L\!\sigma_\mathrm{KN}(\boldsymbol{r}, E)\,n_\mathrm{e}(\boldsymbol{r})\,\mathrm dL\\
  \tau_\mathrm{ph}&=\int_L\!\sum_I\sigma_\mathrm{ph}(\boldsymbol{r},
  E, I)\,n(\boldsymbol{r}, I)\,\mathrm dL\\
  \tau_\mathrm{pp}&=\int_L\!\sum_I\sigma_\mathrm{pp}(\boldsymbol{r},
  E, I)\,n(\boldsymbol{r}, I)\,\mathrm dL,
\end{align}
where $L$ is the path traveled by the photon, $\sigma_\mathrm{KN}$ the
Klein-Nishina cross section, $n_\mathrm{e}$ the electron number
density, $\sigma_\mathrm{ph}$ photoabsorption cross section, $n$ the
element number density, and the sums are taken over all elements $I$.
The path is defined as the line starting from the position of the
previous scattering, or the initial position if the photon has just
been created. This integral is solved continuously as the photon
travels through the ejecta by discretizing the continuous integral
into a sum of finite $\mathrm dL$. The discretization is chosen such
that the distance $\mathrm dL$ is equal to 1\,\% of the magnitude of
the current radial position of the photon. It was verified that finer
discretizations result in similar results and the spatial resolution
is ultimately limited by the resolution of the models (see
Section~\ref{sec:geo}). A photon propagates until $\tau$ reaches a
limiting optical depth $\tau_\mathrm{lim}$, which is sampled from the
distribution \begin{equation}
  \label{eq:tau_pdf}
  \tau_\mathrm{lim} = -\ln\xi_\tau.
\end{equation}
At this point, one of the following three interactions occurs; the
photon scatters off an electron, the photon is photoabsorbed, the
photon pair produces.

\subsection{Interactions}\label{sec:sca_pho}
In this subsection, we describe how the different interactions are
treated numerically at the point of interaction. Scattering affects
the photon energy and the direction of propagation, but does not
destroy the photon; photoabsorption destroys the photon; and pair
production destroys the photon and creates an electron-positron
pair. We assume that no positronium forms and that each positron is
converted into a photon pair locally \citep{chugai97, ruiz_lapuente98,
  jerkstrand11}.

An interaction occurs when $\tau=\tau_\mathrm{lim}$. If the
condition \citep{pozdnyakov83}
\begin{equation}
  \label{eq:ph_vs_cs}
  \xi_\mathrm{ph} < \frac{\tau_\mathrm{ph}}{\tau_\mathrm{cs}+\tau_\mathrm{ph}+\tau_\mathrm{pp}}
\end{equation}
is fulfilled, the photon is absorbed. Absorption means that the photon
is destroyed and the program moves on by simulating the next MC
photon. If Eq.~\eqref{eq:ph_vs_cs} is not fulfilled, but
\begin{equation}
  \label{eq:ph_vs_cs2}
  \xi_\mathrm{ph} < \frac{\tau_\mathrm{ph}+\tau_\mathrm{pp}}{\tau_\mathrm{cs}+\tau_\mathrm{ph}+\tau_\mathrm{pp}}
\end{equation}
is fulfilled, the original photon is replaced by a pair of 511~keV
photons with random directions in the locally comoving frame. This
represents pair production and subsequent annihilation.

If neither condition is fulfilled, scattering occurs and a new
direction and energy are computed.
Both the new photon direction and energy are dependent on the velocity
of the scattering electron. The electron velocities are taken to be
the velocities given by the homologous expansion, implying that all
electrons are moving radially outward. The effect of this compared to
stationary electrons is an increase in escaping number flux of
approximately 2\,\%. We also verified that the thermal motion of
electrons is unimportant compared to the bulk ejecta expansion in the
situation under consideration, where temperatures are lower than
10,000~K.

The next step is to sample a new direction and energy for the photons
after scattering. The distribution of the scattering angle in the
electron rest frame is given by
\citep[e.g.,][]{rybicki79} \begin{equation}
  \label{eq:diff_cs}
  \frac{\mathrm{d}\sigma_\mathrm{KN}^\prime{}}{\mathrm{d}\Omega^\prime{}}\propto
  \left(\frac{E^\prime{}_\mathrm{f}}{E_\mathrm{i}^\prime{}}\right)^2
  \left(\frac{E_\mathrm{i}^\prime{}}{E_\mathrm{f}^\prime{}}+
    \frac{E_\mathrm{f}^\prime{}}{E_\mathrm{i}^\prime{}}-\sin^2\alpha^\prime{}\right),
\end{equation}
where $\Omega$ is the solid angle. We note that this expression
depends on the energy after scattering, which is given by
Eq.~\eqref{eq:frac_ene}. The problem is to sample $\alpha$ from the
distribution obtained by inserting Eq.~\eqref{eq:frac_ene} into
\eqref{eq:diff_cs}. One possibility is to compute the cumulative
distribution function of $\alpha$ numerically and then use the method
of inverse functions \citep[e.g., Section~9.1 of][]{pozdnyakov83} to
draw samples of $\alpha$ from Eq.~\eqref{eq:diff_cs} using uniformly
distributed random numbers. This requires a two-dimensional look-up
table of the cumulative distribution function as a function of
$E_\mathrm{i}$ and $\alpha$. An alternative approach based on a
rejection technique is presented in Section~9.5 of
\citet{pozdnyakov83}, which is also summarized in Appendix~D of
\citet{santana16}\footnote{We note that the plus sign in the
  expression for $\Omega_{3,f}^\prime$ ($\Omega_{z,f}^\prime$ with our
  notation) in \citet{santana16} should be a minus sign.}. We verified
that both methods agree, and choose to implement the rejection
algorithm.

\subsection{Output}
The output of our simulation is a list of photons that escape the
SN. For each photon packet, we save the initial and final energy; the
direction of propagation; the number of scatterings; the time of the
packet as measured by an observer at infinity; and initial SN age when
the packet was created by a radioactive decay. The initial time is
used to weight the packet by the remaining mass of the radioactive
parent nuclei at that time. This weight is equivalent to the number of
photons represented by the packet. We note that we include the
expansion of the ejecta while the photon is traveling. The effect of
including ``live'' expansion of the ejecta is a decrease in escaping
flux of approximately 1\,\%, because while the photon is traveling
outward, the ejecta also expand. The packet lists contain all
information necessary to construct light curves and spectra during
selected time intervals. It is also possible to investigate the light
curves and spectra along different lines of sight.

%%%%%%%%%%%%%%%%%%%%%%%%%%%%%%%%%%%%%%%%%%%%%%%%%%%%%%%%%%%%%%%%
\section{Results}\label{sec:results}
First, in Section~\ref{sec:res_asy}, we present the effects of the
ejecta asymmetries on the escaping X-ray and gamma-ray emission
integrated over all energies, which are important for interpreting the
comparisons with the observed data of \sna{}. Then, we compare the
models that attempt to match \sna{} (B15, N20, \ma{}, \mb{}, and
10HMM) with observational data. Predictions from the B15 and \ma{}
models show best agreement with observations. Therefore, we focus on
these two models and provide more details on their asymmetries. The
spectra are presented in Section~\ref{sec:res_spe}, the continuum
light curves in Section~\ref{sec:res_clc}, and the line fluxes in
Section~\ref{sec:res_lin}. Lastly, we include spectra and line fluxes
for the models that represent other types of SNe in
Section~\ref{sec:res_oth} and investigate the effects of progenitor
metallicity in Section~\ref{sec:res_met}. The line profiles are
investigated in a separate paper (\jerkstrand{}).

Error bars for the observational data correspond to 1$\sigma$
confidence intervals. Exceptions are temporal error bars that
represent the integration periods and spectral bins where the error
bars represent the bin widths.

\subsection{Asymmetries}\label{sec:res_asy}
\begin{figure*}
  \centering
  \includegraphics[width=\columnwidth]{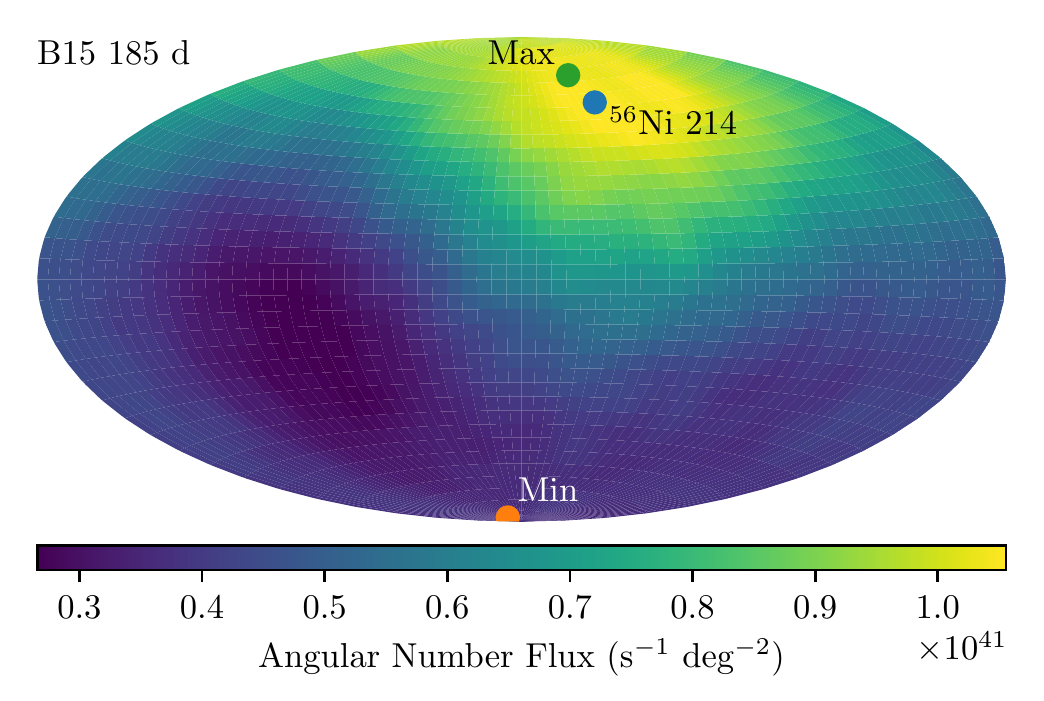}
  \includegraphics[width=\columnwidth]{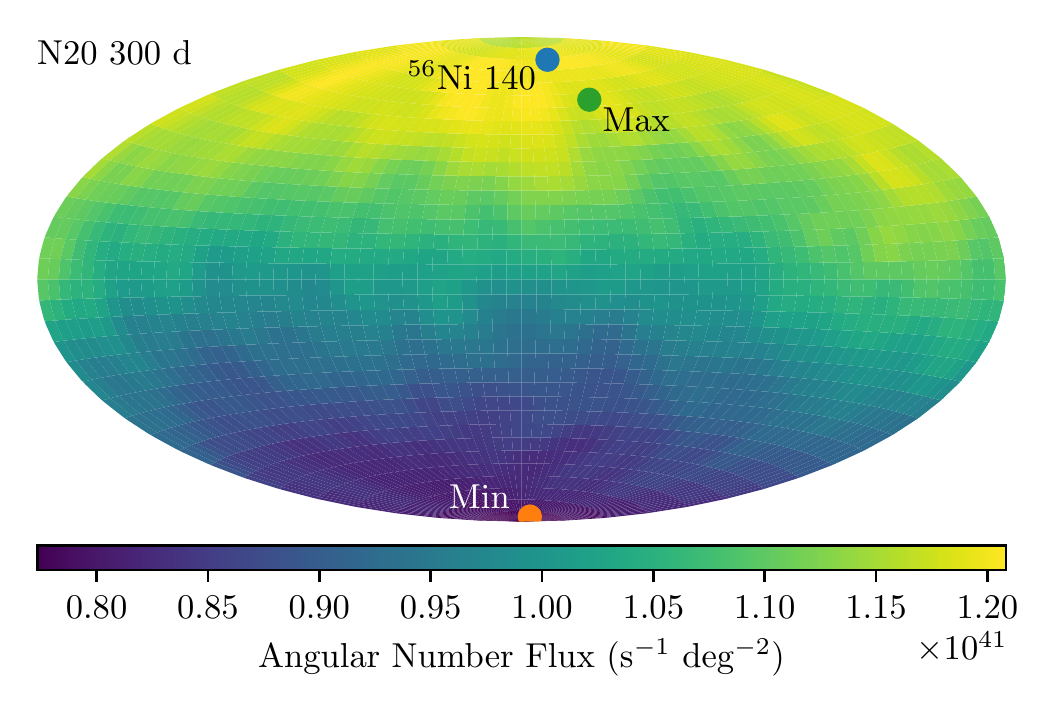}
  \includegraphics[width=\columnwidth]{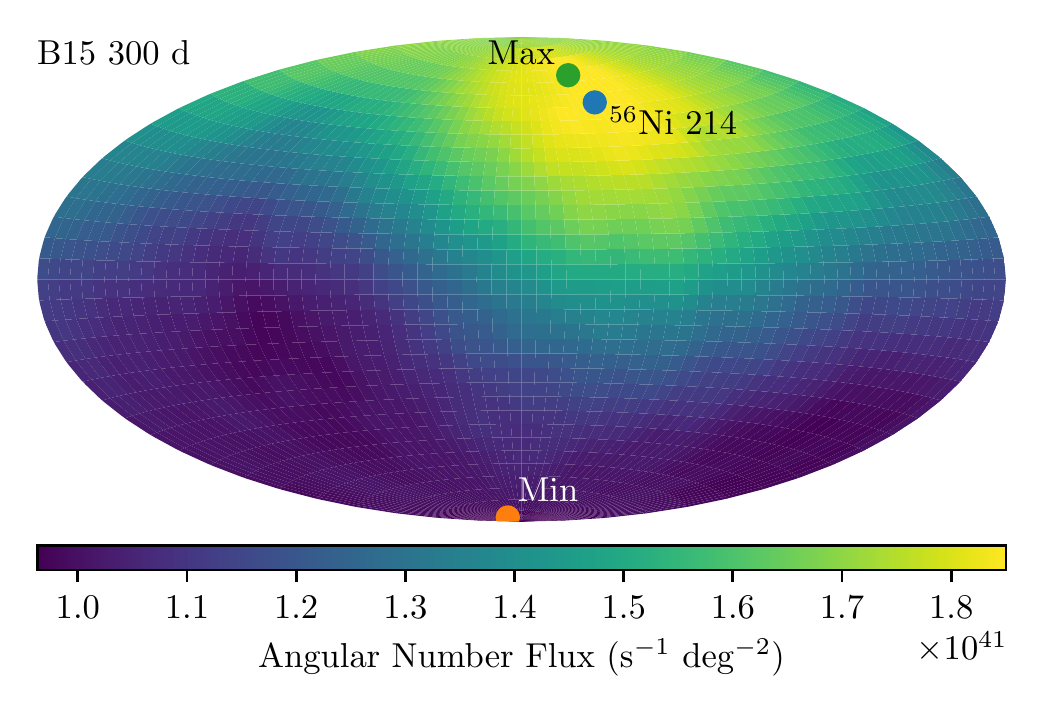}
  \includegraphics[width=\columnwidth]{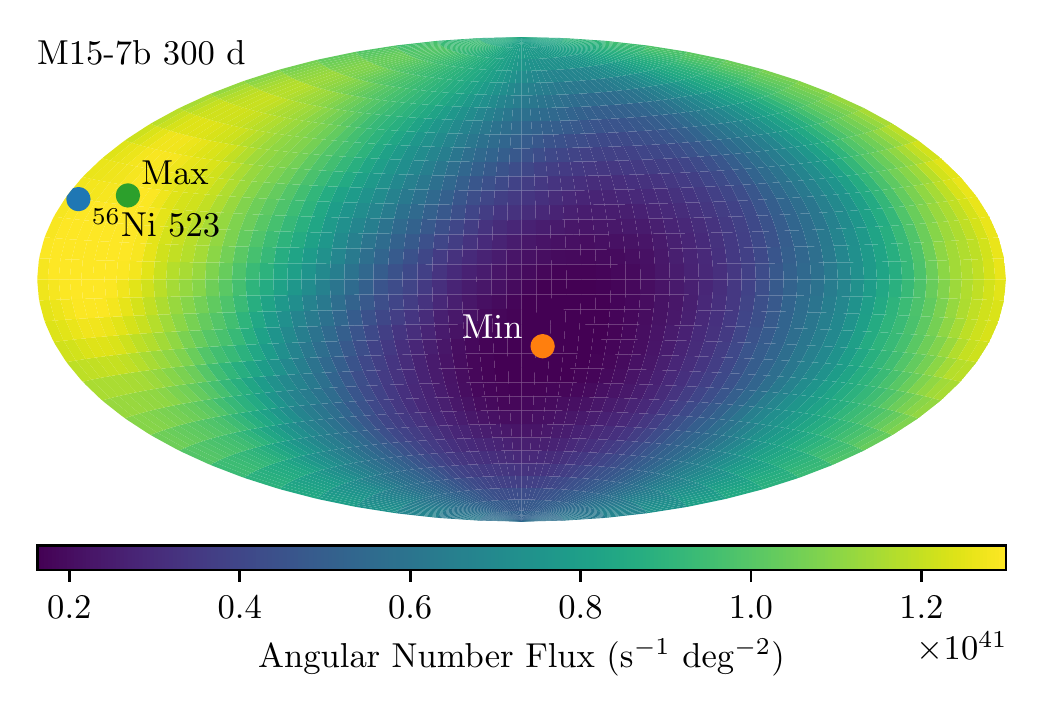}
  \includegraphics[width=\columnwidth]{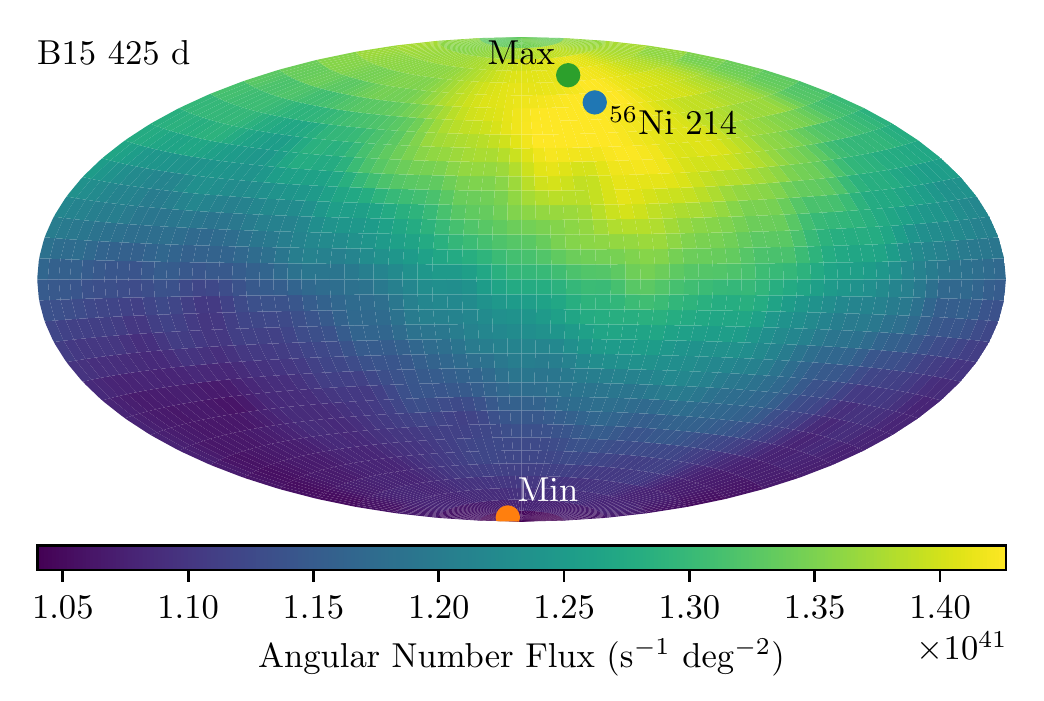}
  \includegraphics[width=\columnwidth]{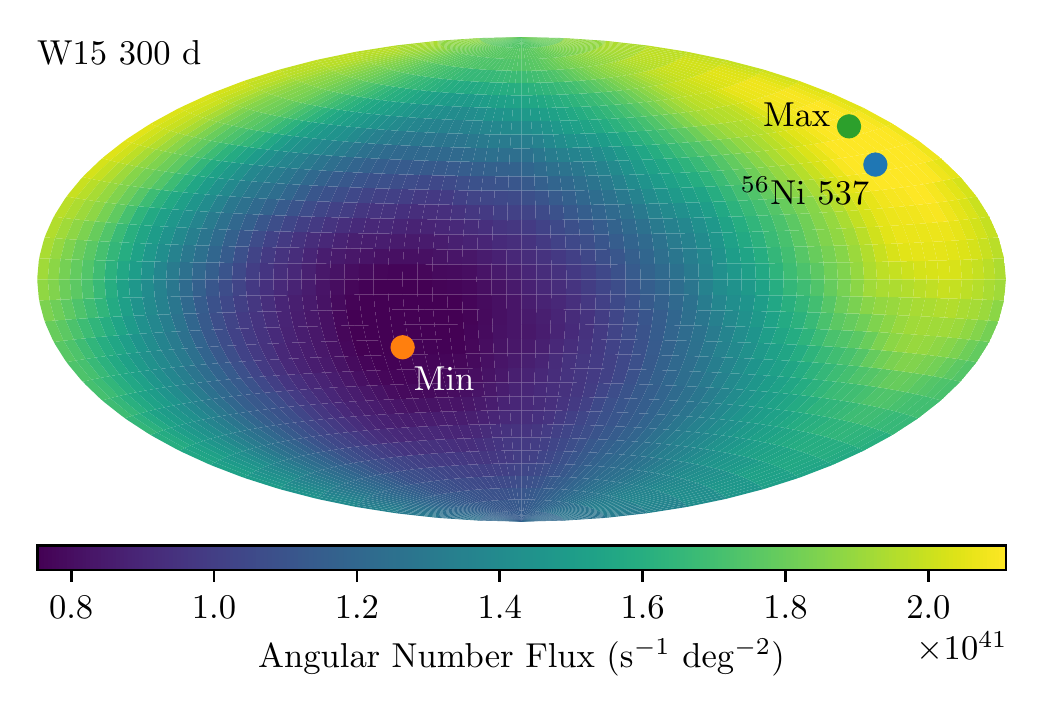}
  \caption{Spherical equal-area Hammer projections of the escaping
    photons at different times (for all energies). The left column
    shows the B15 model at three different times and the right column
    shows the N20, \ma{}, and W15 models at 300~d. The points show the
    direction of the \nc{} center of mass (blue), minimum flux
    (orange), and maximum flux (green). The minimum and maximum
    directions are defined by the extremum number fluxes integrated
    over all energies and times up to 1000~d. The numbers for the
    \nc{} centers of mass are the radial velocities in units
    of\kmps{}. The distributions for narrow energy intervals are
    similar to the integrated distributions. The flux asymmetries are
    initially larger and slowly decrease as the ejecta dilute. The
    main difference is the difference in amplitude of the flux
    asymmetries, which reflects the different levels of ejecta
    asymmetries in the models.\label{fig:sky}}
\end{figure*}
Figure~\ref{fig:sky} shows spherical projections of the escape
directions of the photons for the different models. All models
primarily show variations on large angular scales and the brightness
is correlated with the center of mass of the \nc{}. Generally, the
anisotropies are larger at early times. The temporal evolutions of all
models are relatively similar and projections in narrow energy ranges
show similar features. The main differences in narrow bands are that
the amplitudes of the emission asymmetries are lower for low-energy
photons, which have scattered many times, while the asymmetries are
strongest for the line photons that escape directly
(Sections~\ref{sec:res_spe} and~\ref{sec:res_lin}). The scales in
Figure~\ref{fig:sky} show that the N20 model is the least asymmetric
whereas \ma{} is the most asymmetric model. The models that are not
shown (L15, \mb{}, and \iib{}) show similar emission asymmetry
properties. The flux ratio along the maximum to minimum direction is
${\sim}$1.7 for L15 at 300~d, ${\sim}$4 for \mb{} at 300~d, and
${\sim}$2.1 for \iib{} at 100~d.

In what follows, we investigate quantities averaged over all
directions, as well as along the directions of minimum and maximum
flux. The angle-averaged properties are a good representation of the
distribution of properties over all directions
(Section~\ref{sec:var_asy}). Therefore, the angle-averaged quantities
are useful, although no real observer is able to measure them. The
angle-averaged properties also do not require arbitrary choices of
viewing angles and are less sensitive to MC noise. Because of the
directional variations are negligible on small angular scales, we use
half-opening angles of 30\degr{} when computing quantities along
certain directions. The extremum directions are defined by the
extremum number fluxes integrated over all energies and all times
($< 1000$~d).

\subsection{Spectra}\label{sec:res_spe}
\begin{figure*}
  \centering
  \includegraphics[width=\columnwidth]{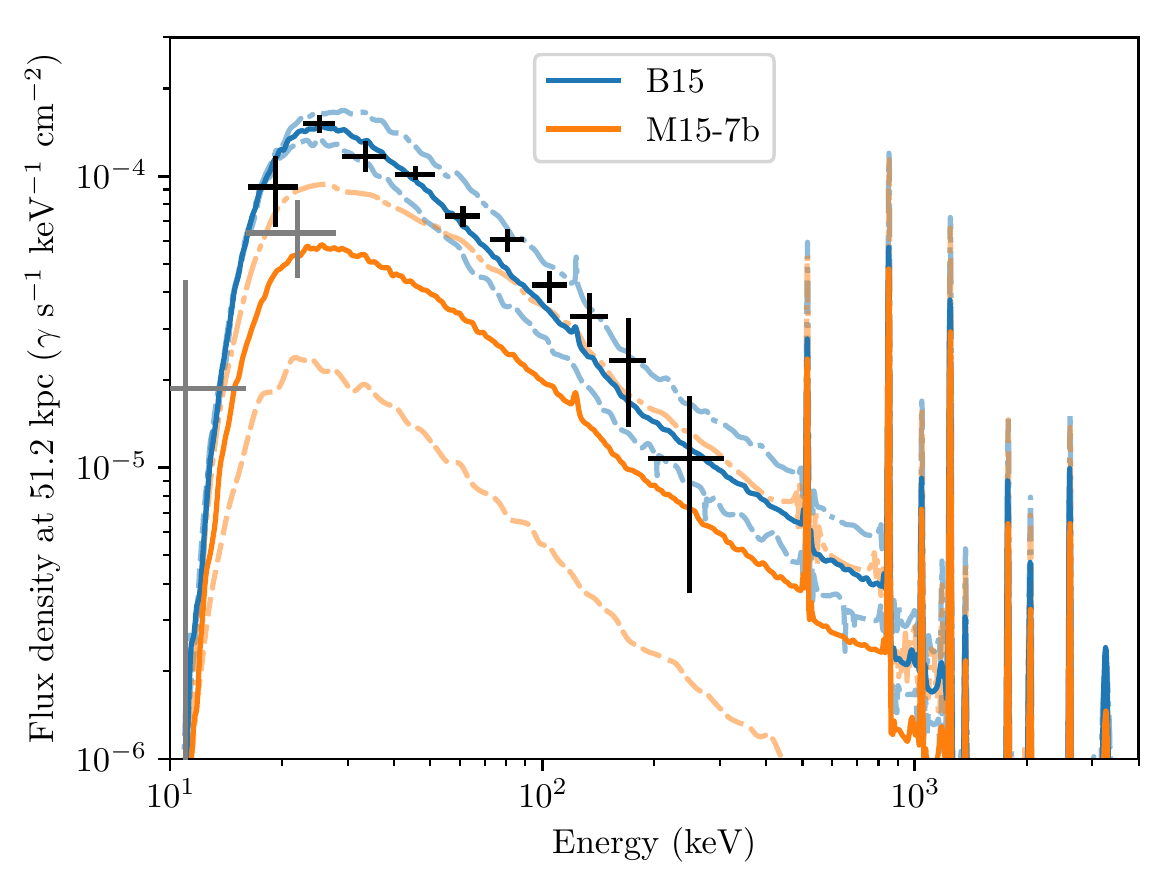}
  \includegraphics[width=\columnwidth]{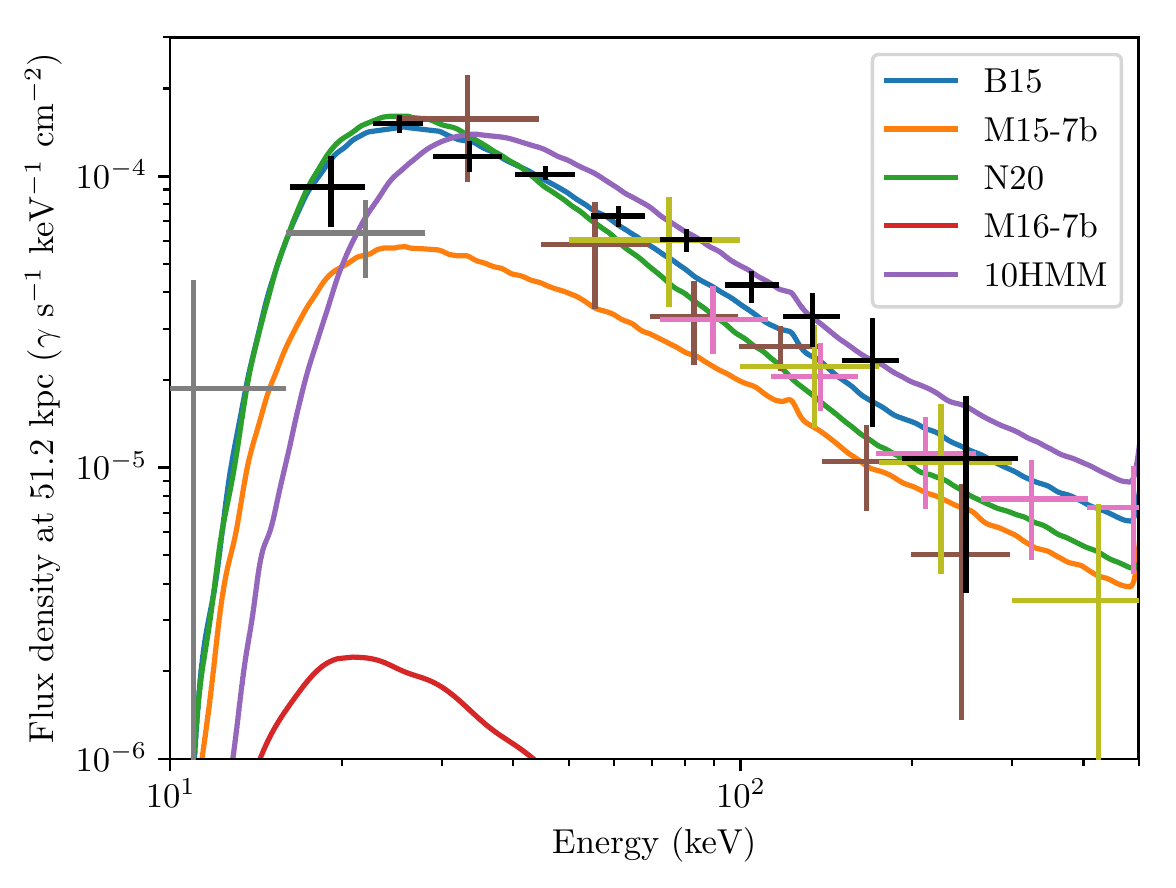}
  \caption{Spectra from the B15 and \ma{} models at 300~d (left) and
    direction-averaged spectra from all \sna{} models
    (right). Observations of \sna{} (crosses); HEXE at 320~d (black,
    \citealt{syunyaev90}), \textit{Ginga} at 300~d (gray,
    \citealt{inoue91}), MSFC at 248~d (brown, \citealt{pendleton95}),
    Pulsar~X\nobreakdash{-}1 at 320~d (pink, \citealt{syunyaev90}),
    and GRIP at 268~d (yellow, \citealt{palmer93}). In the left panel,
    the solid lines are the spectra averaged over all directions, the
    semi-transparent dashed lines are along the directions of minimum
    flux, and the semi-transparent dash-dotted lines are along the
    directions of maximum flux. The envelope metallicity, which sets
    the low-energy cutoff, of the 10HMM model is not well-defined
    because metals were artificially mixed into parts of the envelope
    (Sections~\ref{sec:mod}
    and~\ref{sec:res_met}) \label{fig:spec_b15}}
\end{figure*}
The left panel of Figure~\ref{fig:spec_b15} shows the averaged and
directional spectra for the B15 and \ma{} models at 300~d and the
corresponding observational data. At 300~d, both the shapes and the
amplitudes are in reasonable agreement with the data and any remaining
deviations between models and observations can relatively easily be
accommodated by the variance introduced by the asymmetries. An
important property is that the asymmetries change the overall
normalizations, but does not affect the shapes much. The typical
amplitude of the flux variations for different viewing angles spans a
factor of ${\sim}$2, but the most asymmetric model, \ma{}, shows
variations up to a factor of ${\sim}$5.

The right panel of Figure~\ref{fig:spec_b15} compares all spectra of
the \sna{} models with observational data around 300~d. We use the
direction-averaged spectra and simply note that variations of a factor
of a few in amplitude are possible because of asymmetries, in
particular at early times (Section~\ref{sec:var_asy}). The N20 model
matches the observed spectra well but is only doing so around 300~d
(Section~\ref{sec:res_clc}). The 1D model 10HMM also agrees relatively
well with the observations, but this required mixing introduced by hand
(Section~\ref{sec:mod}). This shows that 3D models are able to
self-consistently produce mixing at levels similar to what is inferred
from observations, whereas this had to be introduced artificially in
1D models.

The \mb{} model (red line toward the bottom of
Figure~\ref{fig:spec_b15} right) clearly fails to match the
observations. We reiterate (Section~\ref{sec:mod}) that \ma{} and
\mb{} are presented here because they fit the X-ray and gamma-ray
observations best and worst, respectively, out of the six merger
progenitors that fulfill the observational criteria of the SN1987A
progenitor \citep{menon17}. Out of the remaining four merger models
(not shown), M15\nobreakdash-8b and M17\nobreakdash-7a are very
similar to \mb{}, whereas M16\nobreakdash-4a and M17\nobreakdash-8b
are intermediate cases. Only \ma{} agrees reasonably well with the
X-ray and gamma-ray observations, whereas the remaining five merger
models can be ruled out.

\subsection{Continuum Light Curves}\label{sec:res_clc}
\begin{figure*}
  \centering
  \includegraphics[width=\columnwidth]{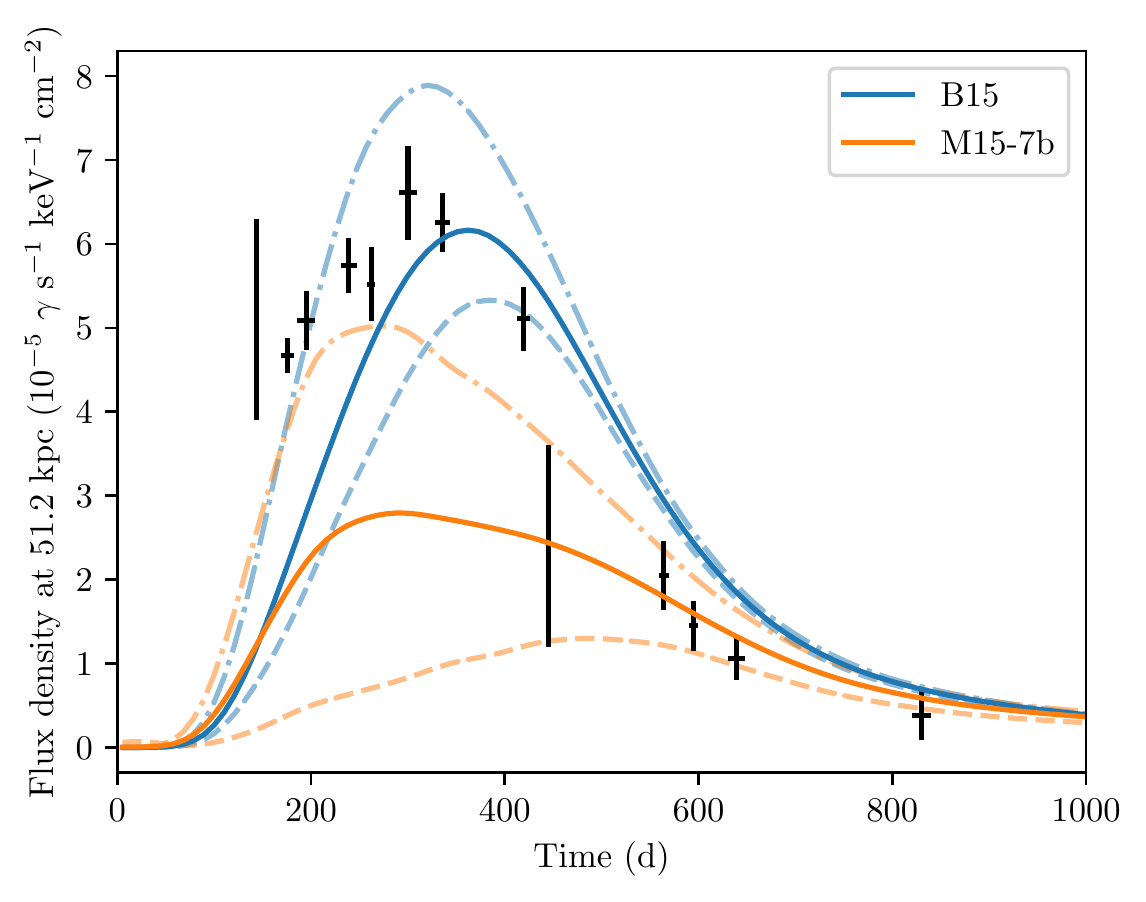}
  \includegraphics[width=\columnwidth]{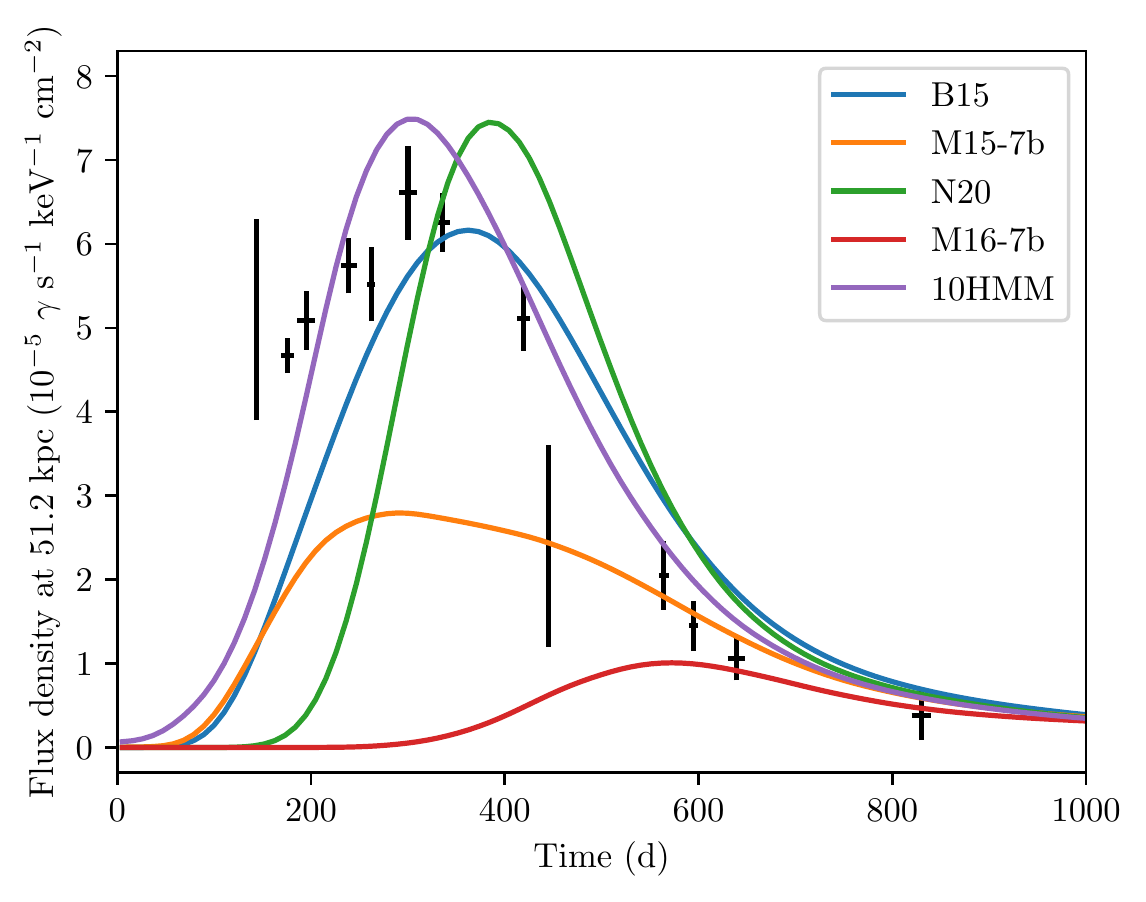}
  \caption{Light curves in the 45--105~keV range from the B15 and
    \ma{} models (left), direction-averaged 45--105~keV light curves
    from all \sna{} models (right), and HEXE \sna{} observations
    (black crosses). The horizontal bars of the data points at 145~d
    and 445~d are not visible because of the short exposures. In the
    left panel, the solid lines are averaged over all directions, the
    semi-transparent dashed lines are along the directions of minimum
    flux, and the semi-transparent dash-dotted lines are along the
    directions of maximum flux.\label{fig:lc_b15}}
\end{figure*}
The light curves in different energy bands show similar results, but
we focus on the 45--105~keV observations by HEXE because they are the
most accurate. The left panel of Figure~\ref{fig:lc_b15} shows the
averaged and directional light curves for the B15 and \ma{} models in
the 45--105~keV range and the corresponding HEXE data. The asymmetries
clearly affect the flux magnitude and the time of the initial rise. In
contrast, the declining tails are relatively similar along different
directions. This is a manifestation of the emission asymmetries
becoming less pronounced at later times.

The average 45--105~keV light curves for all models and the HEXE
observations are shown in the right panel of Figure~\ref{fig:lc_b15}.
None of the models is able to match the early observed breakout time
and all overshoot the light curve at later times (except for \mb{},
which completely fails to match the observations). The general
agreement with observations, however, can still be considered
acceptable given the uncertainties and sensitivity of the emission
properties on the progenitor structure
(Section~\ref{sec:constraints}).

\subsection{Line Emission}\label{sec:res_lin}
\begin{figure*}
  \centering
  \includegraphics[width=\columnwidth]{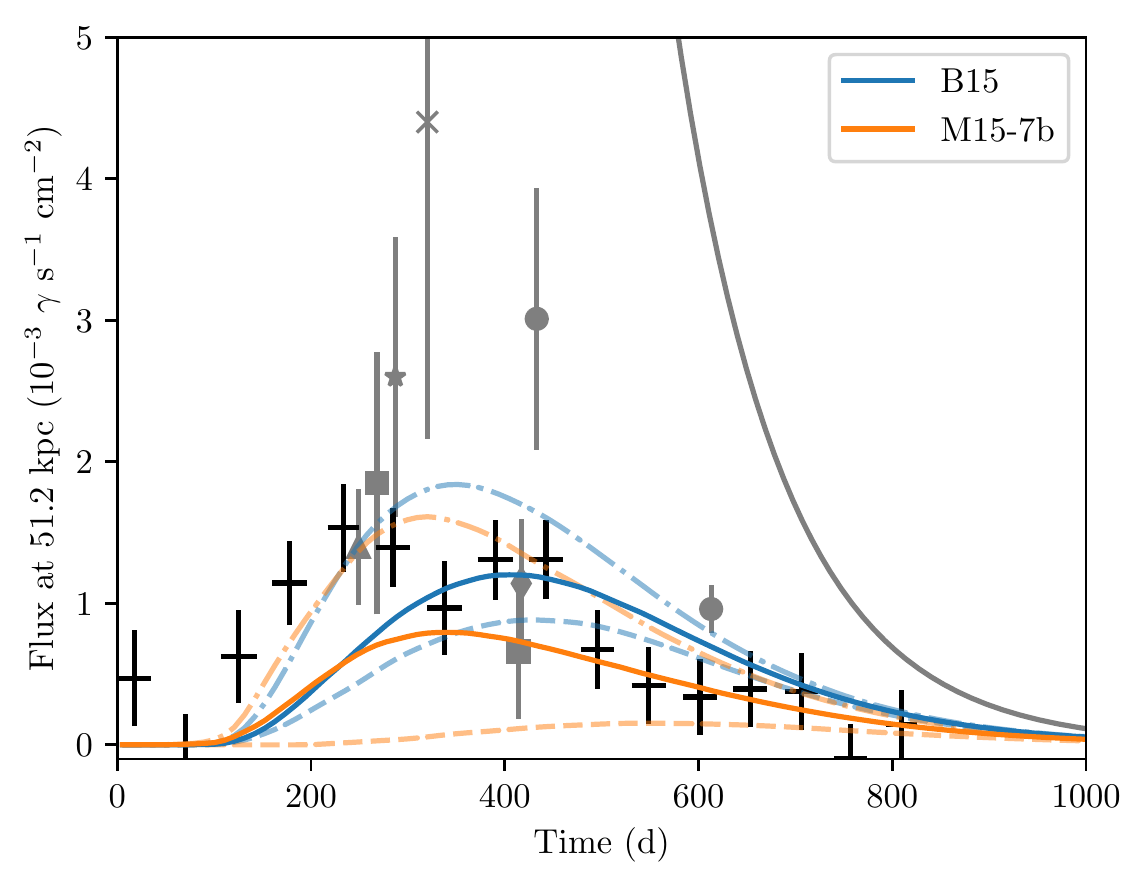}
  \includegraphics[width=\columnwidth]{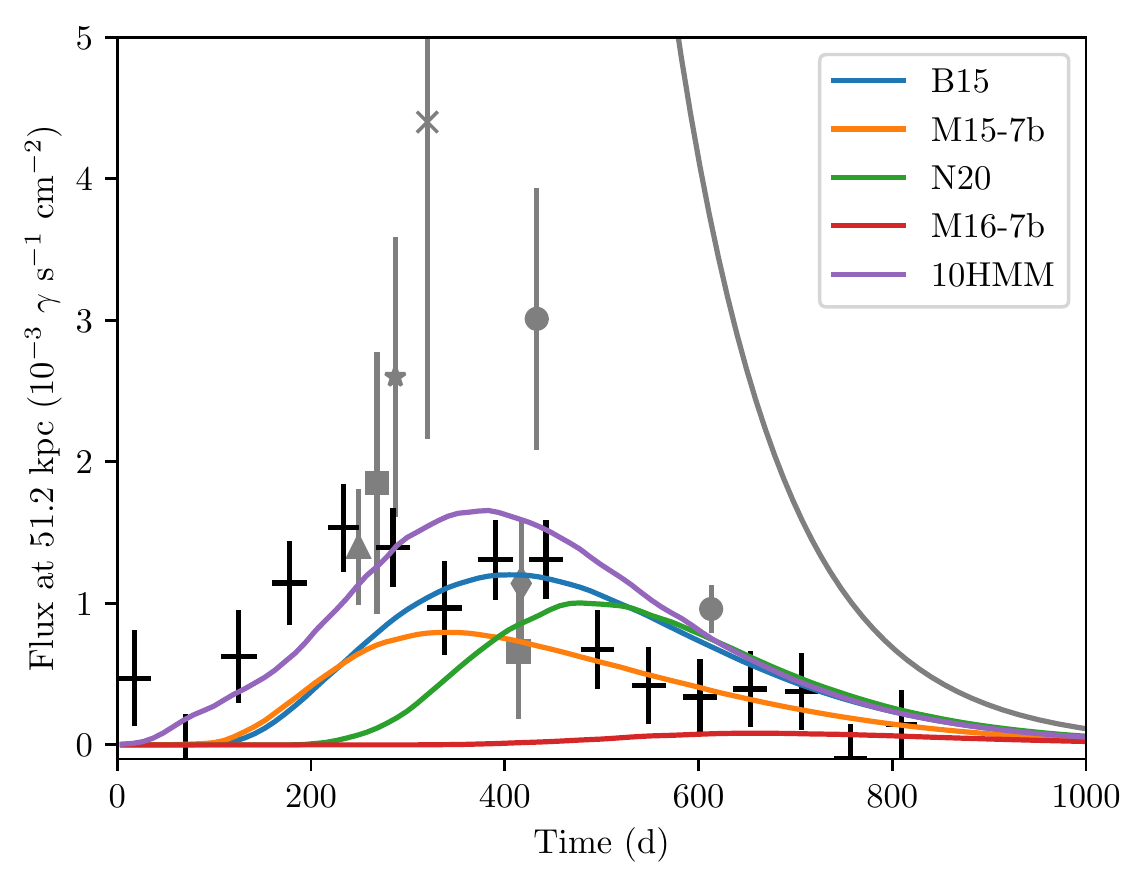}
  \caption{Temporal evolution of the sum of the 847 and 1238~keV line
    fluxes from the B15 and \ma{} models (left) and direction-averaged
    fluxes from all \sna{} models (right). The colored solid lines
    that are included in the legend are direction-averaged model
    predictions and the gray curve is the analytic limit of free
    escape for 0.07\Msun{} of initial \nc{} \citep{suntzeff90,
      bouchet91}. The semi-transparent dashed lines are along the
    minimum flux directions and the dash-dotted are along the maximum
    flux directions of the B15 and \ma{} models. The black crosses are
    \textit{SMM} measurements. The gray markers are the balloon
    measurements; MSFC (triangle, \citealt{sandie88}), GRIP (squares,
    \citealt{palmer93}), GRIS (circles, \citealt{tueller91b}), JPL
    (star, \citealt{mahoney88}), GRAD ($\times{}$,
    \citealt{rester89}), and UCR (diamond,
    \citealt{ait_ouamer92}).\label{fig:llc}}
\end{figure*}
Figure~\ref{fig:llc} shows the line fluxes of the sum of the 847 and
1238~keV lines for the \sna{}-like models. For the observations that
only covered one of the lines, we scale that value by the atomic line
yields to obtain the expected combined flux under the assumption of
equal optical depths at 847 and 1238~keV. There is possibly a slight
indication that the \textit{SMM} measures lower fluxes than the
balloon-borne experiments \citep{leising91, teegarden94}. For the
comparisons with model predictions, we primarily focus on the
\textit{SMM} values because of the continuous coverage and homogeneity
of the data.

The predicted line fluxes show the same breakout time problem as the
continuum light curves. The line flux is more sensitive to the viewing
angle than the continuum emission in the sense that the relative
differences in flux are larger. This is because the continuum photons
have scattered into new directions before escaping the ejecta, which
effectively reduces the strength of the asymmetries. Apart from this,
the viewing angle also affects the light curve shape for the more
asymmetric models, similarly as for the continuum light curves. The
accuracy of the observed line data is not as good as the continuum
precision. However, the line fluxes are only functions of the optical
depth along the photon path, which makes them valuable for breaking
degeneracies when interpreting the more complex continuum data
(Section~\ref{sec:constraints}).

\subsection{Other SN Types}\label{sec:res_oth}
\begin{figure}
  \centering
  \includegraphics[width=\columnwidth]{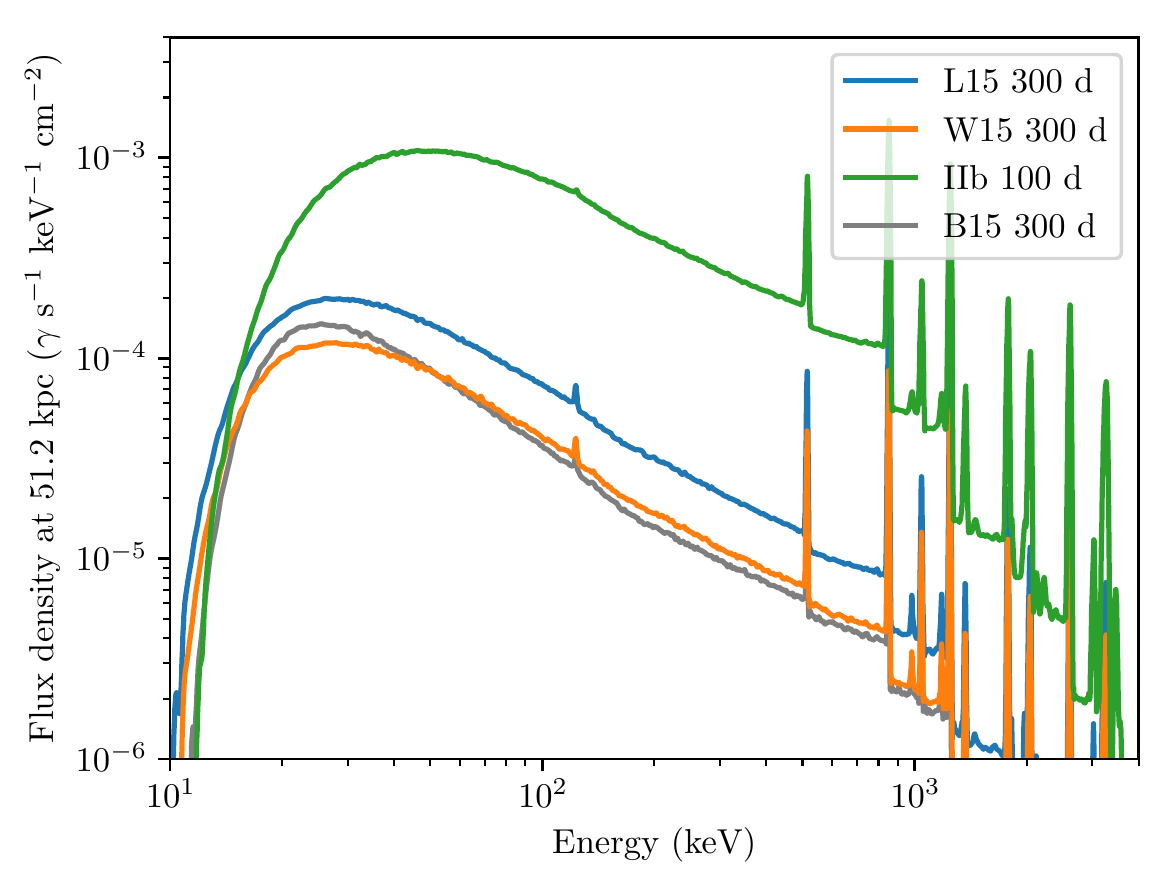}
  \caption{Direction-averaged spectra of the RSG models L15 and W15 at
    300~d, and the stripped IIb model at 100~d. The gray line is the
    B15 model at 300~d and is included for reference. The times are
    chosen to be around peak flux. The variations along different
    directions primarily affect the magnitude of the flux while the
    shape remains relatively constant (Figure~\ref{fig:spec_b15}). The
    differences at the low-energy cutoff are primarily attributed to
    the different metallicities (Table~\ref{tab:mod} and
    Section~\ref{sec:res_met}).\label{fig:oth_spe}}
\end{figure}
We include predictions for other types of SNe as a guide for future
observations. We reiterate that the masses of the radioactive elements
in all models are scaled to the inferred values of \sna{}
(Section~\ref{sec:nuc_net}). Figure~\ref{fig:oth_spe} shows the
spectra of the RSG models L15 and W15, as well as the stripped IIb
model. We note that the \hrich{} models are shown at 300~d, whereas
the \iib{} model is at 100~d. This roughly corresponds to the times of
peak flux. The spectral shape for a given model is softer at earlier
times and harder at later times. We note that the RSG models are
similar to the B15 model.

\begin{figure}
  \centering
  \includegraphics[width=\columnwidth]{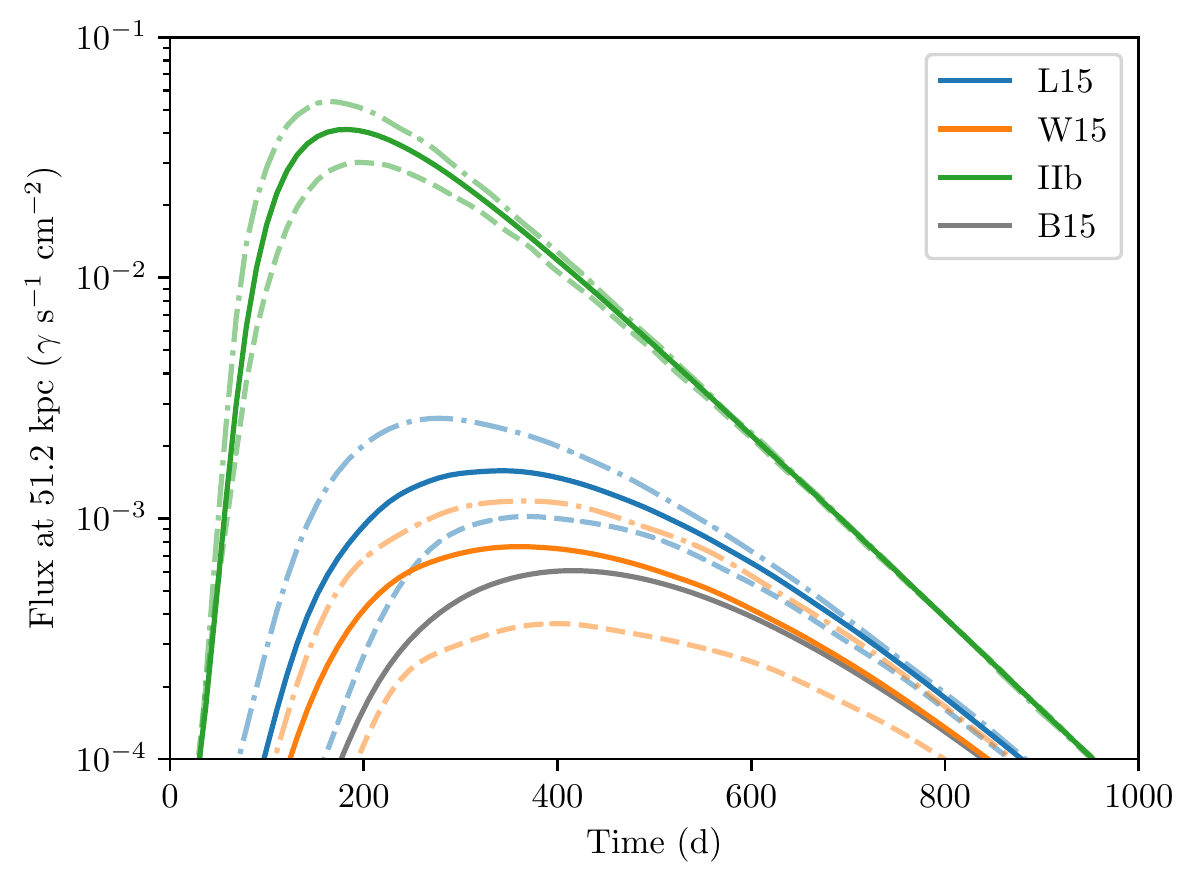}
  \caption{Temporal evolution of the 847~keV line fluxes (not the sum
    of the 847 and 1238~keV lines as in Figure~\ref{fig:llc}) of the
    RSG models L15 and W15, and the stripped IIb model. The gray line
    is the direction-averaged line flux from the B15 model, which is
    included for reference. The semi-transparent dashed colored lines
    represent the flux along the minimum direction and the
    semi-transparent dash-dotted lines the flux along the maximum
    direction for the corresponding models.\label{fig:oth_lc}}
\end{figure}
Figure~\ref{fig:oth_lc} shows the 847~keV line fluxes of the L15, W15,
and IIb models. We also include the light curves along the extremum
directions. The variations due to asymmetries are flux variations by a
factor of a few but the qualitative properties of the line fluxes are
independent of the viewing angle.

The most notable features of the IIb model are that it evolves faster
and is more luminous at the time of peak flux than the \hrich{}
models. This is because of the lower ejecta mass and higher expansion
velocities. Furthermore, the very thin H envelope quickly becomes
transparent, which shifts the photoabsorption cutoff to slightly
higher energies, because the core is revealed and a larger fraction of
the photons escape the ejecta before being scattered many times.

\subsection{Effects of Progenitor Metallicity}\label{sec:res_met}
In this subsection, we use five versions of the B15 model to
investigate the effects of different progenitor envelope
metallicities. In addition to the B15 version with LMC abundances
(0.55\Zeffsun{}), we generate three additional versions of B15 with
effective metallicities of 0.12, 0.27, and 1.00\Zeffsun{}. These three
versions are corrected using solar abundances (Table~\ref{tab:abu})
following the method presented in
Section~\ref{sec:progenitor_metallicity}. Lastly, we create a fifth
version with abundances corresponding to the ring of \sna{}
(Section~\ref{sec:dis_pro_met}), which results in an effective
metallicity of 0.28\Zeffsun{}.

\begin{figure}
  \centering
  \includegraphics[width=\columnwidth]{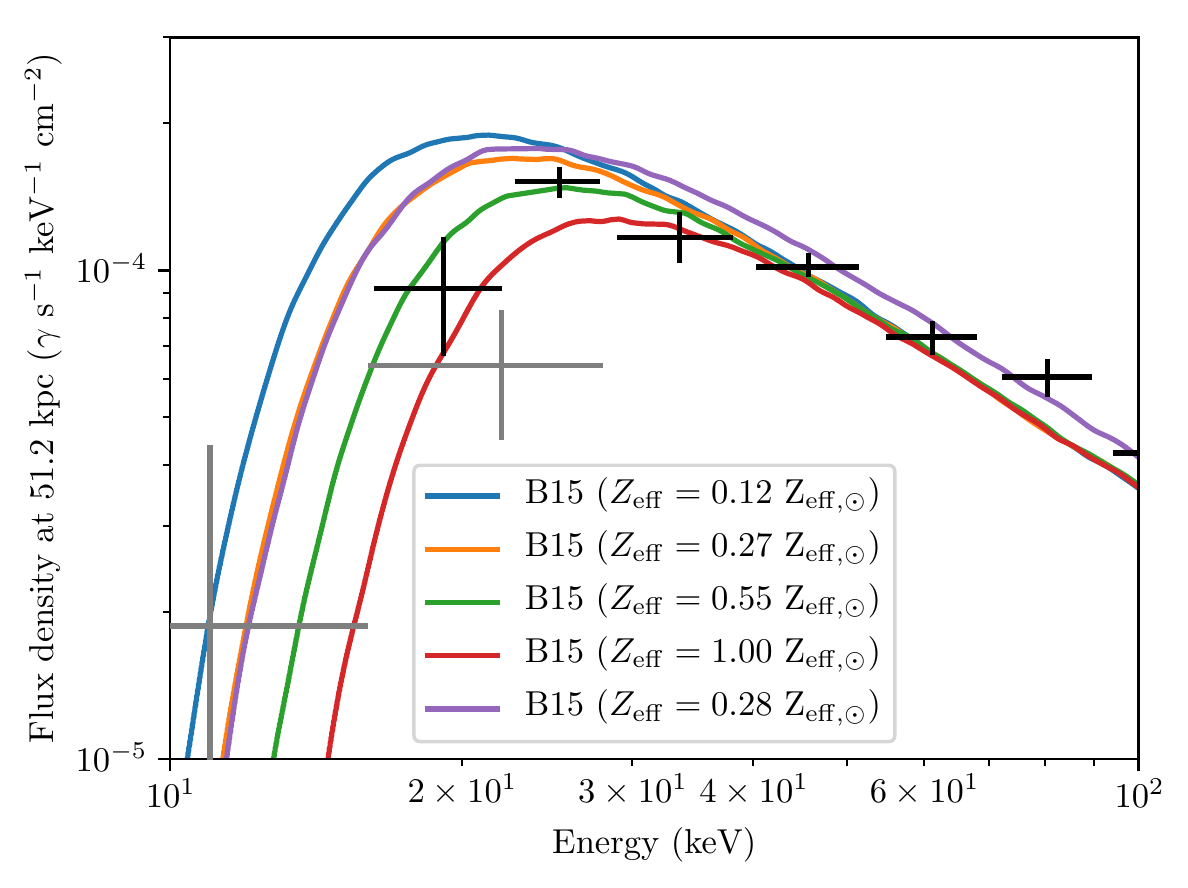}
  \caption{Spectra at 300~d for the B15 model at five different
    metallicities. This shows how increasing metallicity (primarily Fe
    abundance) of the progenitor envelope affects the low-energy
    photoabsorption cutoff. The green 0.55\Zeffsun{} line is for LMC
    abundances and the purple 0.28\Zeffsun{} line is for abundances of
    the ring of \sna{}. The purple line deviates at higher energies
    because of a much higher He-to-H ratio
    (Section~\ref{sec:dis_pro_met}), which changes the Compton
    scattering opacity. Overplotted are the observed HEXE spectrum at
    320~d (black crosses), and the \textit{Ginga} bands at 300~d (gray
    crosses).\label{fig:zz}}
\end{figure}
Figure~\ref{fig:zz} shows spectra at 300~d for the five different B15
models. This clearly shows how increasing metallicity shifts the
low-energy cutoff of the spectra to higher energies by increasing the
photoelectric absorption opacity. The observed \sna{} spectra seem to
align particularly well with the version with LMC metallicity, but is
also consistent with abundances inferred from the ring of SN 1987A.
The four versions that are practically identical at higher energies
have almost the same Compton scattering opacity. This is because the
metallicity corrections only involve minor changes in terms of mass,
which only marginally affect the scattering opacity. The reason why
the version with the \sna{} ring abundances differs at higher energies
is because of a major shift of 2.1\Msun{} of H into He
(Section~\ref{sec:dis_pro_met}). This reduces the electron density and
significantly decreases the scattering opacity. However, the cutoff is
still at an energy similar to that of the version with 0.27\Zeffsun{},
which was corrected using solar abundances.

%%%%%%%%%%%%%%%%%%%%%%%%%%%%%%%%%%%%%%%%%%%%%%%%%%%%%%%%%%%%%%%%
\section{Discussion}\label{sec:discussion}
\subsection{General Model Emission Properties}\label{sec:gen}
Before comparing the predictions with the \sna{} observations
(Section~\ref{sec:constraints}), we discuss some general properties of
the predicted emission. The X-ray and gamma-ray emission from
different progenitors shows different properties. For a given viewing
angle, the most important parameter is the electron column density
outside of the fastest \nc{} or, in other words, the amount of mixing
(Section~\ref{sec:mod}). This determines the time at which the
first emission starts escaping the ejecta. An additional major
difference between the models is that the stripped IIb model evolves
on shorter timescales and is more luminous than the \hrich{} models
because of its lower ejecta mass and higher explosion velocities.

\subsubsection{Variance Due to Asymmetries}\label{sec:var_asy}
\begin{figure*}
  \centering
  \includegraphics[width=\columnwidth]{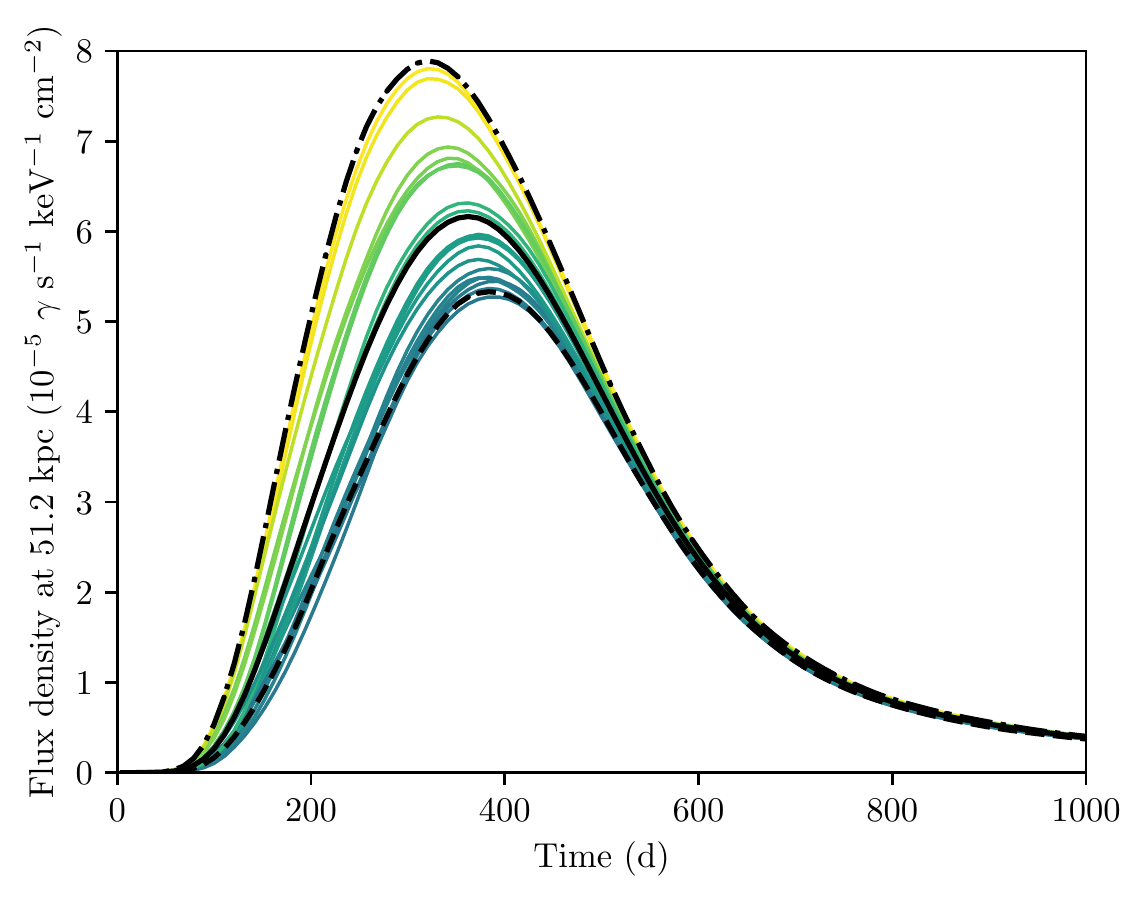}
  \includegraphics[width=\columnwidth]{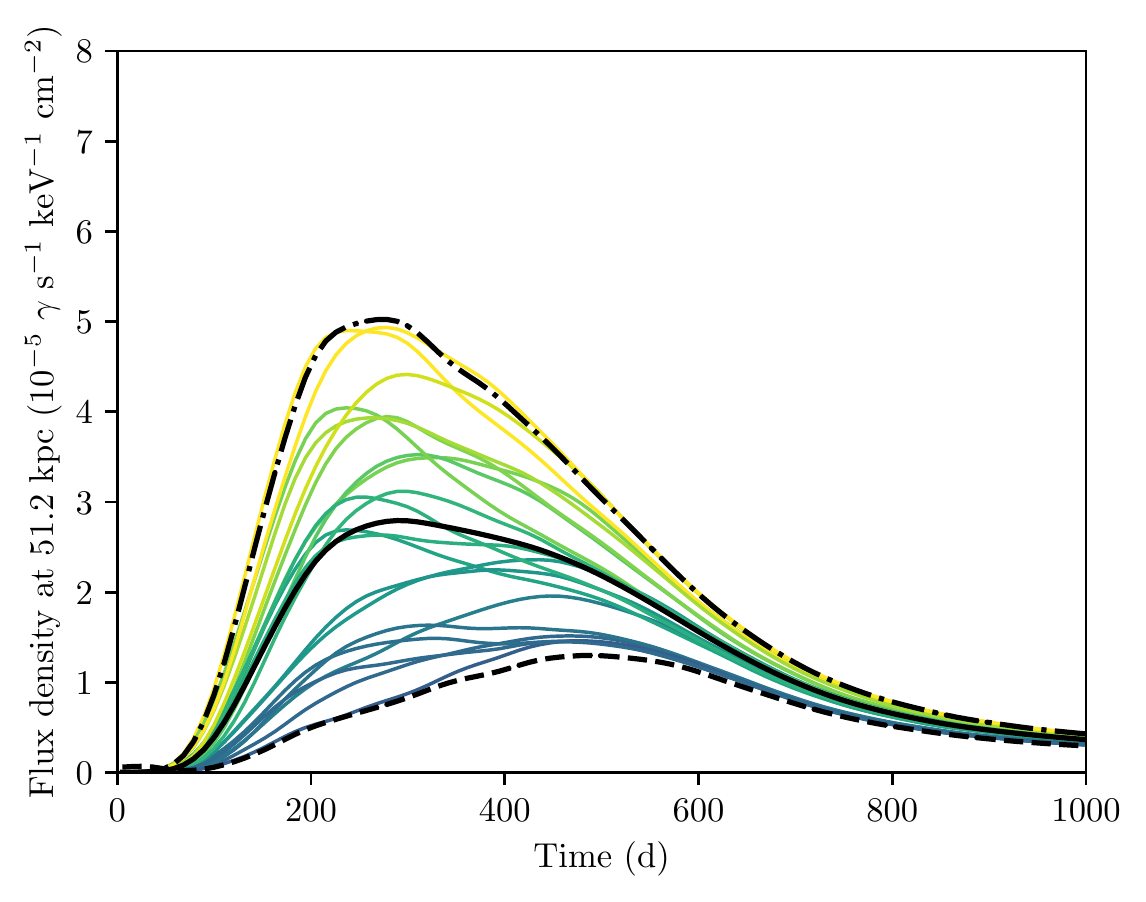}
  \caption{Light curves in the 45--105~keV range along different
    viewing angles for the B15 model (left) and the \ma{} model
    (right).  The solid black lines are averaged over all directions,
    the dashed black lines are along the directions of minimum flux,
    and the dash-dotted black lines are along the directions of
    maximum flux.  The colored lines are along 20 arbitrary uniformly
    distributed directions (the vertices of a
    dodecahedron). \label{fig:var_asy}}
\end{figure*}
First, we explore the effects of the choice of viewing angle on the
observed properties. This is the variance introduced by the 3D
structures for a given explosion. The variance arising from the
stochastic hydrodynamic instabilities for repeated explosions of the
same progenitor is investigated in Section~\ref{sec:var_hyd}.

We show the light curves in the 45--105~keV range for the B15 and
\ma{} models along different viewing angles in
Figure~\ref{fig:var_asy}. The effects of the asymmetries are overall
changes in amplitudes and shapes of the light curves. This is more
prominent for the \ma{} model, which is the most asymmetric model. The
shapes of the spectra, however, are only weakly affected
(Figure~\ref{fig:spec_b15}). The general behavior of the spectral
shapes along different directions is larger differences at higher
energies, particularly for the direct line emission, than at lower
energies. This is because the scattering effectively smooths out the
asymmetries. This also implies that the line fluxes are more strongly
affected by asymmetries than the 45--105~keV light curves. The peak
line flux varies along different directions by more than a factor of 5
in B15 and 10 in \ma{}, and the times of peak line flux differ by up
to 100~d in B15 and 300~d in \ma{} between the minimum and maximum
directions.

Finally, Figure~\ref{fig:var_asy} clearly shows that the range of
fluxes spanned by the fluxes along the minimum and maximum directions
contains the fluxes from practically all directions at all times. It
also shows that the angle average is a good representation of the
distribution of properties over all directions.

\subsubsection{Variance Due to Explosion Dynamics}\label{sec:var_hyd}
\begin{figure*}
  \centering
  \includegraphics[width=\columnwidth]{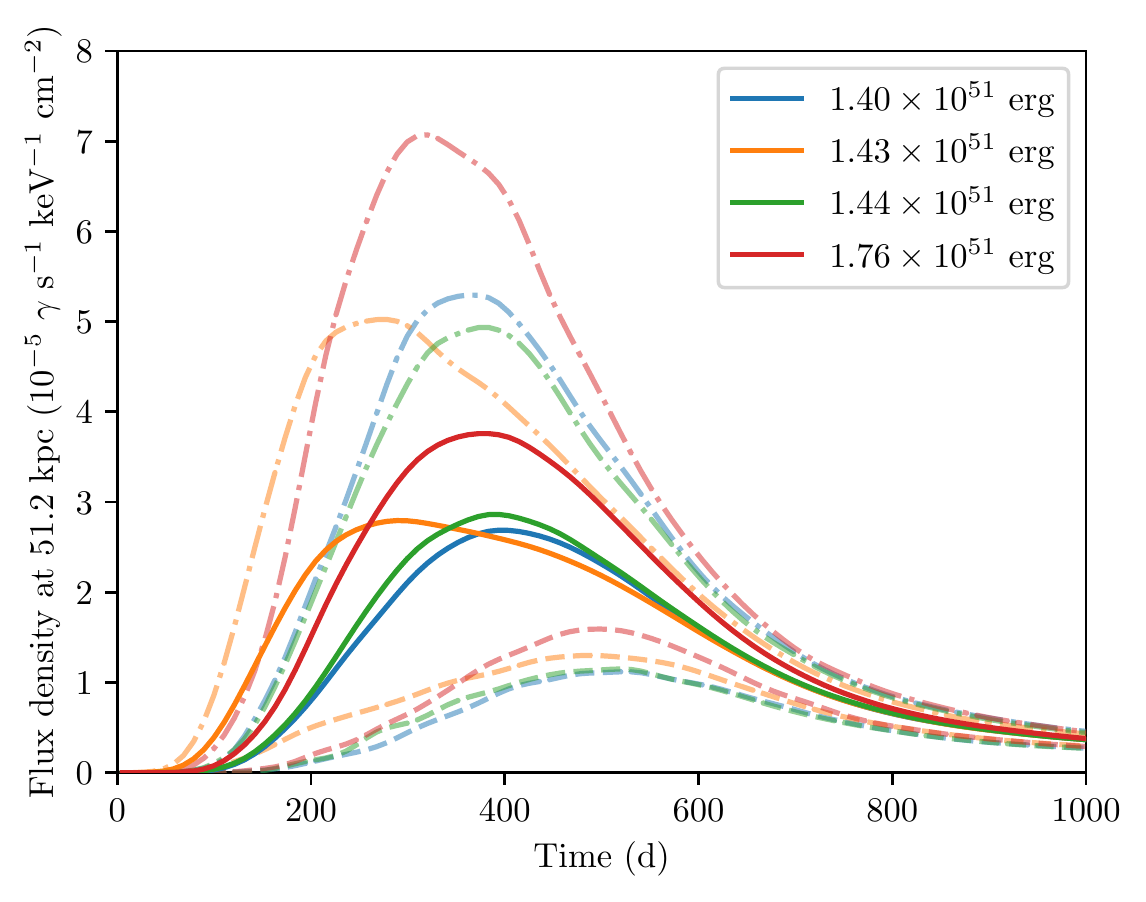}
  \includegraphics[width=\columnwidth]{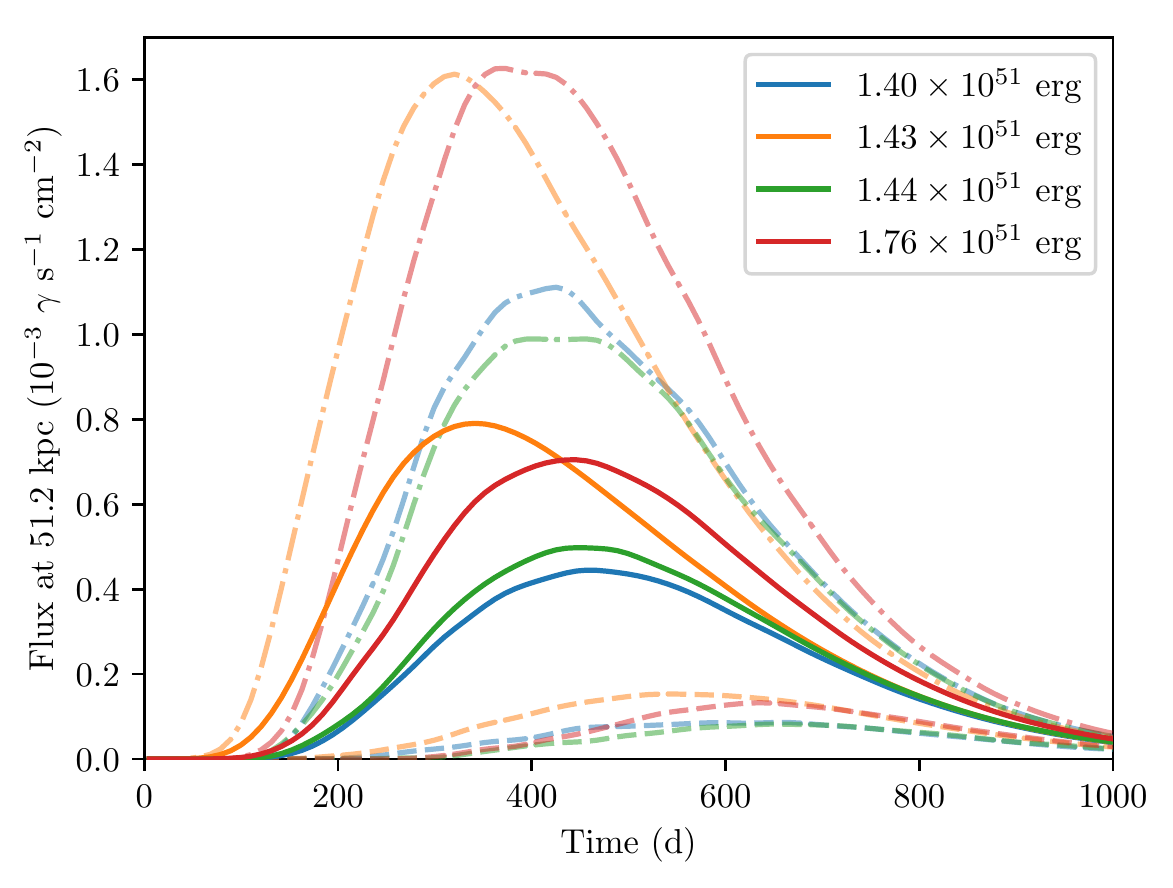}
  \caption{Light curves in the 45--105~keV range (left) and the sum of
    the 847 and 1238~keV line fluxes (right) from four versions of the
    \ma{} model with different explosion energies and seed
    perturbations. The orange lines are the reference \ma{}, which we
    focus on throughout the rest of the paper. The solid lines are
    averaged over all directions, the semi-transparent dashed lines
    are along the directions of minimum flux, and the semi-transparent
    dash-dotted lines are along the directions of maximum
    flux. \label{fig:var_dyn}}
\end{figure*}
The variance introduced by the stochastic hydrodynamic evolution is
seeded by the random fluctuations in the structures of the progenitors
\citep{wongwathanarat13}. We investigate the effects on the emission
properties by simulating three additional explosions of the \ma{}
model. The first explosion differs by having different seed
perturbations in the mapping from the 1D progenitor into three
dimensions. The final explosion energy of this second model is
$1.44\times{}10^{51}$~erg. The two other explosions were simulated
with different seed perturbations and slightly different neutrino
luminosities. They result in explosion energies of
$1.40\times{}10^{51}$ and $1.76\times{}10^{51}$~erg. A similar
exercise was carried out using optical light curves for three versions
of the B15 model by \citet{utrobin15}.

The 45--105~keV light curves for the four versions of \ma{} are shown
in Figure~\ref{fig:var_dyn} (left). In addition to the variance due to
asymmetries of the individual models, the peak flux of the 45--105~keV
continuum varies within ${\sim}$30\,\% and the peak time shifts by up
to ${\sim}$100~d. This means that the light curve shapes are slightly
different. The line fluxes (Figure~\ref{fig:var_dyn}, right) show
similar variance as the continua, with the primary difference being
that the angle-averaged line flux peaks span a factor of 2. For \ma{},
the differences in spectral shape resulting from the stochastic nature
of the explosion are much smaller than the variance due to the
intrinsic asymmetries. Finally, B15 and \ma{} can still be
distinguished, despite the broad distributions of properties due to
both the stochasticity of the explosions and intrinsic asymmetries
(Figures~\ref{fig:var_asy} and~\ref{fig:var_dyn}).

\subsubsection{Progenitor Metallicity}\label{sec:dis_pro_met}
The progenitor surface metallicity also has interesting implications
for the X-ray properties. Increasing progenitor metallicity shifts the
low-energy X-ray cutoff to higher energies (Figure~\ref{fig:zz}). This
effect could potentially be used to constrain the progenitor
metallicity and has previously been discussed in the context of
Type~Ia SNe \citep{maeda12}. The cutoff is also not very sensitive to
the viewing angle (Figure~\ref{fig:spec_b15}) because it is determined
by the properties of the outermost layers of the ejecta, which are
more isotropic. On the other hand, if the homogeneity of the outer
envelope is broken by large-scale convection (e.g.,
\citealt{hoflich91}), it is likely that the low-energy cutoff would be
less sharp, which would make it more difficult to constrain the
metallicity. Inversely, this could also serve as a diagnostic of the
envelope isotropy.

It is important to point out that the cutoff is only dependent on the
envelope metallicity when the low-energy limit is set by
photoabsorption. This is the case in the early phases during the X-ray
rise. At times later than the X-ray flux peak, the escaping emission
is dominated by radiation from the deeper parts where the mean atomic
number is higher and photoelectric cross sections are larger. Another
contributing factor (especially at very late times) is that the
low-energy cutoff is determined by the ability of the ejecta to trap
photons, which determines how many times photons scatter and lose
energy before they escape.

\begin{figure}
  \centering
  \includegraphics[width=\columnwidth]{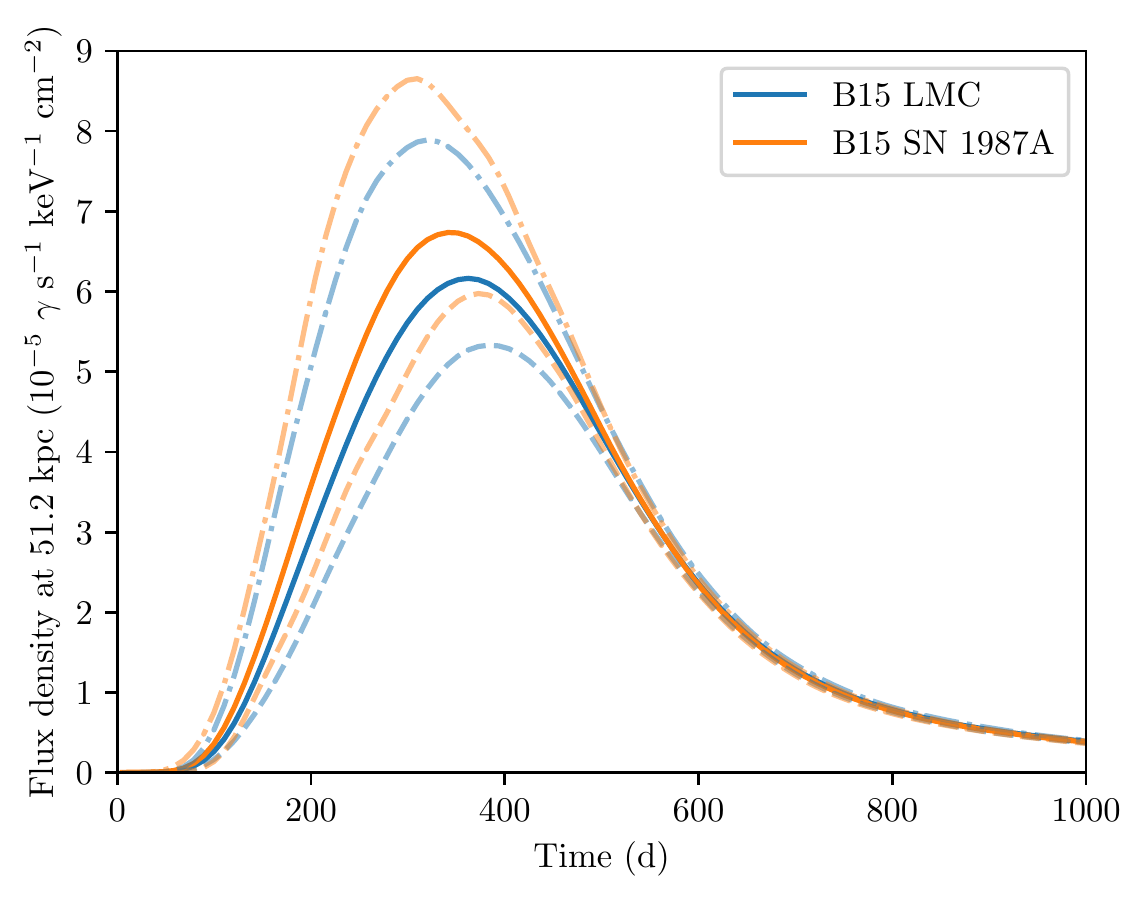}
  \caption{Light curves in the 45--105~keV range for the standard B15
    version with LMC abundances (blue) and a modified B15 version with
    \sna{} abundances (orange). The solid lines are averaged over all
    directions, the semi-transparent dashed lines are along the
    directions of minimum flux, and the semi-transparent dash-dotted
    lines are along the directions of maximum flux. \label{fig:abu}}
\end{figure}
In Section~\ref{sec:progenitor_metallicity}, we describe the
metallicity corrections for the B15 and N20 models to LMC abundances
(Table~\ref{tab:abu}). An alternative is to use the abundances of the
\sna{} progenitor inferred from observations of the equatorial
ring. We check the effects on the results by comparing the B15 model
with LMC abundances (that is used throughout the rest of the paper) to
a B15 version with \sna{} abundances. The 45--105~keV light curves of
these models are shown in Figure~\ref{fig:abu}. The line fluxes behave
in a similar way. The differences are primarily driven by the increase
of the of He abundance to 11.31 (0.21 relative to H by number;
\citealt{lundqvist96, mattila10}) and the decrease of the Fe abundance
to 6.98 ($9.6\times{}10^{-6}$ relative to H by number,
\citealt{dewey08, zhekov09, mattila10, dewey12}).  Both changes
contribute to lowering the effective metallicity to 0.28\Zeffsun{}
and, consequently, shift the photoabsorption cutoff to lower energies
(Figure~\ref{fig:zz})\footnote{The conversion of H to He does not
  directly lower the effective metallicity (in fact, the effective
  metallicity increases slightly). However, this indirectly lowers the
  abundances of the metals, which dominate the photoabsorption
  opacity, because we adopt values of all elements relative to
  H. Thus, the net effect is a decrease of the effective
  metallicity.}. The inferred \sna{} progenitor He abundance is more
than twice as high as the LMC abundance (Table~\ref{tab:abu}). This
requires a conversion of 2.1\Msun{} of H into He in the envelope,
which is a much more drastic change to the progenitor than the
correction to LMC abundances. The H-to-He change lowers the scattering
opacity because of the lower number of electrons per unit mass, which
in turn leads to an earlier time of rise and a higher peak
flux. Overall, the differences are smaller than the intrinsic
asymmetries of the B15 model.

\subsection{Comparisons with \sna{}}\label{sec:constraints}
The predictions of the B15 and \ma{} models (along the maximum
direction) capture the general features of the observed data
relatively well. There are, however, some significant differences that
have constraining implications for the models. The other \sna{} models
are worse at reproducing the observations, primarily because of
insufficient mixing of the \nc{} to the outer layers
(Table~\ref{tab:mod}). The amount of mixing depends on the properties
of the progenitor in a complex way and cannot be directly inferred
from basic progenitor or explosion parameters
(Section~\ref{sec:mod}). Understanding this complex relation requires
a more detailed analysis and evaluation of the growth rates of
Rayleigh-Taylor instabilities for each of the models. This has been
investigated for the H-rich single-star progenitors
\citep{wongwathanarat15, utrobin19}, and both single stars and mergers
will be presented in a forthcoming paper (\utrobin{}). We do not
discuss N20, \mb{}, or the other merger models (not presented) further
since they all fail to match the observations.

Focusing on B15 and \ma{}, we investigate what can be inferred from
the remaining differences between the model predictions and the
observations. The property that is easiest to interpret is the line
flux. At early times during the rising phase, this practically only
depends on the column density of electrons outside the fastest trace
amounts of \nc{} on the near side. In contrast, at later times when
the line fluxes are declining, they are practically only a function of
the average absorption through the ejecta to the bulk of the \nc{}. It
is clear from Figure~\ref{fig:llc} that the line fluxes of all models
fail to capture the early observed rise before 200~d and the fast
decline after 400~d. The early-time observations most likely imply
that trace amounts of \nc{} were ejected toward us at slightly higher
velocities (more strictly, higher mass coordinate) than what is seen
in the models. The only other (less likely) option is that there is a
thinner ``hole'' through the envelope that allows some emission from
deeper regions to escape at early times. The late-time line
observations imply that there is more material that absorbs the direct
emission than predicted by the models. This can be achieved by either
larger total ejecta masses or the \nc{} being preferentially ejected
toward the far side away from us.

The line fluxes are closely related to the continuum light curves. The
main difference is that the continuum light curves depend
non-trivially on the optical depths to the radioactive elements. For
example, the continuum emission is quenched for very high absorption,
as well as when approaching the optically thin regime, because the
continuum requires down-scattering of line photons. In conjunction
with the observations of the direct line emission, however, it is
straightforward to break this degeneracy. From
Figure~\ref{fig:lc_b15}, it is clear that the predicted 45--105~keV
light curves fail to reach the early observed fluxes before 200~d and
overshoot the observed values at times later than 400~d, similarly to
the line fluxes. This results in the same constraints on the
distribution of \nc{} as discussed above, but is still helpful because
the continuum data are more accurate and also provide additional
independent observations.

The model spectra at 300~d in Figure~\ref{fig:spec_b15} agree very
well with observations. However, the same early deficits and late
excesses that are seen in the continuum light curves
(Figure~\ref{fig:lc_b15}) are of course also present in the spectra at
early and late times (not shown). The difference between predicted
spectra and observations at these times is primarily a change in
normalization, which also implies that the continuum light curves in
other energy bands show similar trends as in the presented 45--105~keV
range. A notable feature in the spectra is the low-energy cutoff. It
can be seen in the right panel of Figure~\ref{fig:spec_b15} that the
\sna{} models match the spectral break around 20~keV. This is simply a
manifestation of the metallicities of the progenitor envelopes
(Section~\ref{sec:res_met}), which are consistent with the observed
X-ray cutoff.

Finally, we stress two points concerning the magnitude of the
discrepancies between the predictions and the observations. First,
even though the rise of the predicted light curves are too late (left
panels of Figures~\ref{fig:lc_b15} and~\ref{fig:llc}), the relative
difference in maximum \nc{} velocity required to match data is
relatively low. We find that artificially increasing the radial
velocity of all \nc{} by around 20\,\% in the B15 model is sufficient
for the direction-averaged emission to match both the low-energy
continuum and line flux rise.

Secondly, the difference by a factor of ${\sim}$2 in the direct line
flux around 600~d (Figure~\ref{fig:llc}) can be remedied by shifting
the \nc{} center of mass. The relevant quantity is the effective
optical depth and, by comparing with the free-escape asymptote in
Figure~\ref{fig:llc}, it is clear that only a slight increase in the
optical depth is sufficient for models to match data. We make another
toy model by taking the original B15 model and moving the \nc{} center
of mass from 214 to 514\kmps{} along the same direction. This is done
by applying a constant shift to all \nc{}, effectively moving the
distribution as a rigid body within the rest of the ejecta. This
results in a relatively good match with observations at late times. It
is only meaningful to view this model from the minimum direction
because a natural consequence of the modification is that the opposite
direction matches the data worse. The increased radial velocity and
the center-of-mass shift of the \nc{} distribution only marginally
affect the spectral shape. It is also worth pointing out that the
aforementioned example of increased mixing and center-of-mass shift is
only one of many possibilities to match the data due to the large
freedom when modifying 3D structures by hand.

\subsection{Future Observations}
We make simple predictions for observations of future nearby SNe by
comparing our results with the sensitivity of current
telescopes.
The \textit{Chandra X-Ray Observatory} (\citealt{weisskopf00,
  weisskopf02, garmire03}) covers soft X-rays below ${\sim}10$~keV,
\textit{NuSTAR} covers the 3--79~keV range \citep{harrison13,
  madsen15}, and the spectrometer SPI \citep{vedrenne03} on board the
\textit{International Gamma Ray Astrophysics Laboratory}
(\textit{INTEGRAL}, \citealt{winkler03}) extends from 20~keV to
8~MeV. Even though the effective area of \textit{XMM-Newton}
\citep{jansen01,struder01,turner01} is larger than the effective area
of \textit{Chandra}, their point source sensitivities are similar
(Figure~6 of \citealt{takahashi10}). For reference, we also include
the sensitivity curve of \astrogam{} \citep{de_angelis17}, which was a
candidate mission for the ESA M5 call and was proposed to operate from
300~keV to 3~GeV.

For the predictions, we choose a specific set of three non-stripped
models and the stripped-envelope IIb model. The B15 version used for
these predictions is without the metallicity correction described in
Section~\ref{sec:progenitor_metallicity}, which means that its
effective metallicity is 0.03\Zeffsun{}. The W15 model is modified to
$Z_\mathrm{eff} = 1.12$\Zeffsun{} using solar abundances. In contrast,
the metallicity of the L15 and the IIb models are unmodified from
their standard values provided in Table~\ref{tab:mod}. We construct
this set of models to illustrate the effects of different
metallicities because of its importance for the low-energy cutoff. The
effects of the metallicity on the direct line fluxes are negligible.

\begin{figure}
  \centering
  \includegraphics[width=\columnwidth]{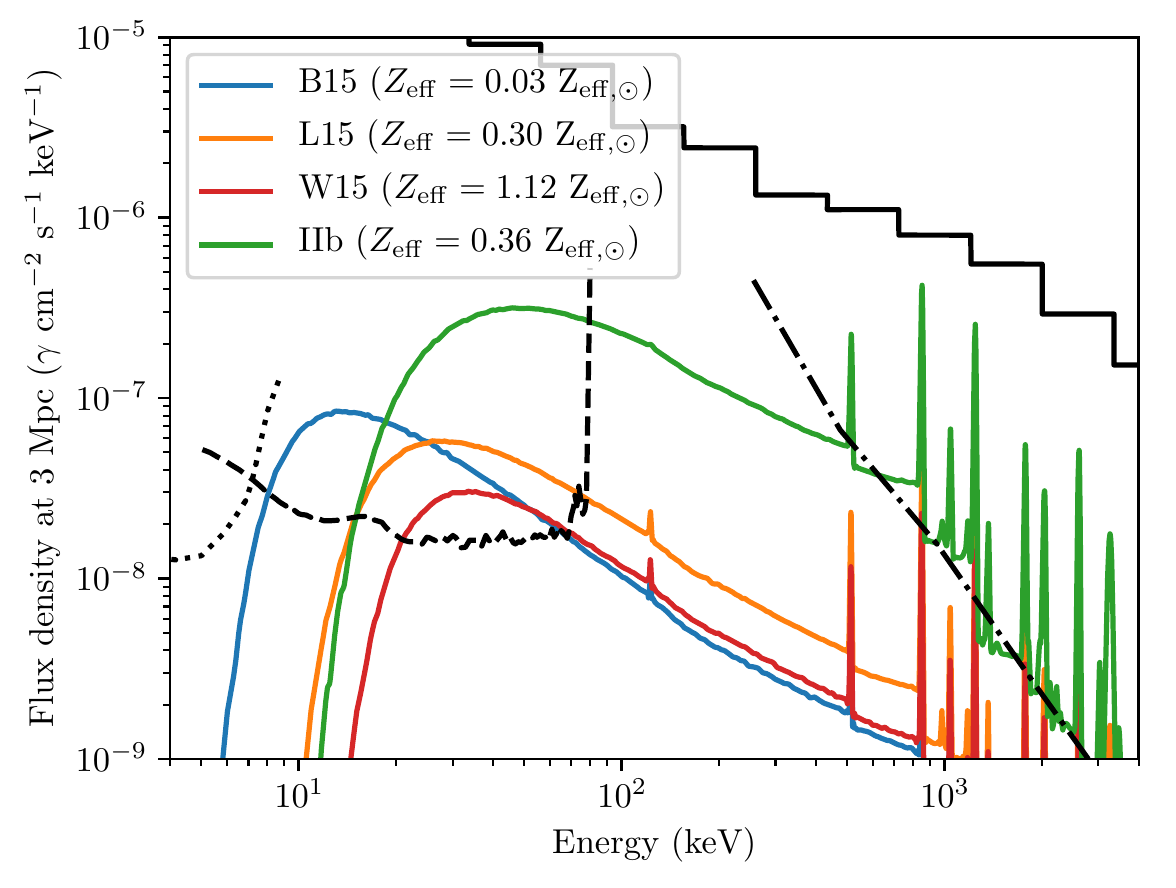}
  \caption{Comparison of predicted continua scaled to 3~Mpc with the
    detection sensitivities of four instruments: \textit{Chandra}
    (dotted black line, \citealt{takahashi10}), \textit{NuSTAR}
    (dashed black line, \citealt{koglin05}), \textit{INTEGRAL} (solid
    black line, \citealt{roques03}), and the ESA M5 proposal
    \astrogam{} (dash-dotted black line, \citealt{de_angelis17}). The
    spectra are at a time of 300~d for the non-stripped models and 100
    d for the IIb model, which are approximately the times of peak
    flux in the \textit{NuSTAR} band. All continuum sensitivities are
    given for spectral bins of $\Delta E/E=0.5$, a detection threshold
    of 3$\sigma$, and an exposure time of
    1~Ms.\label{fig:sensitivity_continuum}}
\end{figure}
Figure~\ref{fig:sensitivity_continuum} shows predicted spectra
overplotted on the sensitivity curves of the
instruments. \textit{NuSTAR} is expected to provide the deepest
observations. The non-stripped models are relatively similar and are
expected to be detectable by \textit{NuSTAR} to around 3~Mpc, whereas
the limiting distance for the IIb model is around 10~Mpc. This is in
agreement with the value of ${\sim}$4~Mpc given by \citet{harrison13}
for \mbox{CCSNe} in general. These distances extend to slightly beyond
the Local Group. It is worth pointing out that the low-metallicity
version of B15 has the photoabsorption cutoff above the
\textit{Chandra} range (our code does not include the much fainter
bremsstrahlung component at lower energies,
Section~\ref{sec:methods}). This means that even metal-free
progenitors do not extend into the soft X-ray regime $<10$~keV.

\begin{figure}
  \centering
  \includegraphics[width=\columnwidth]{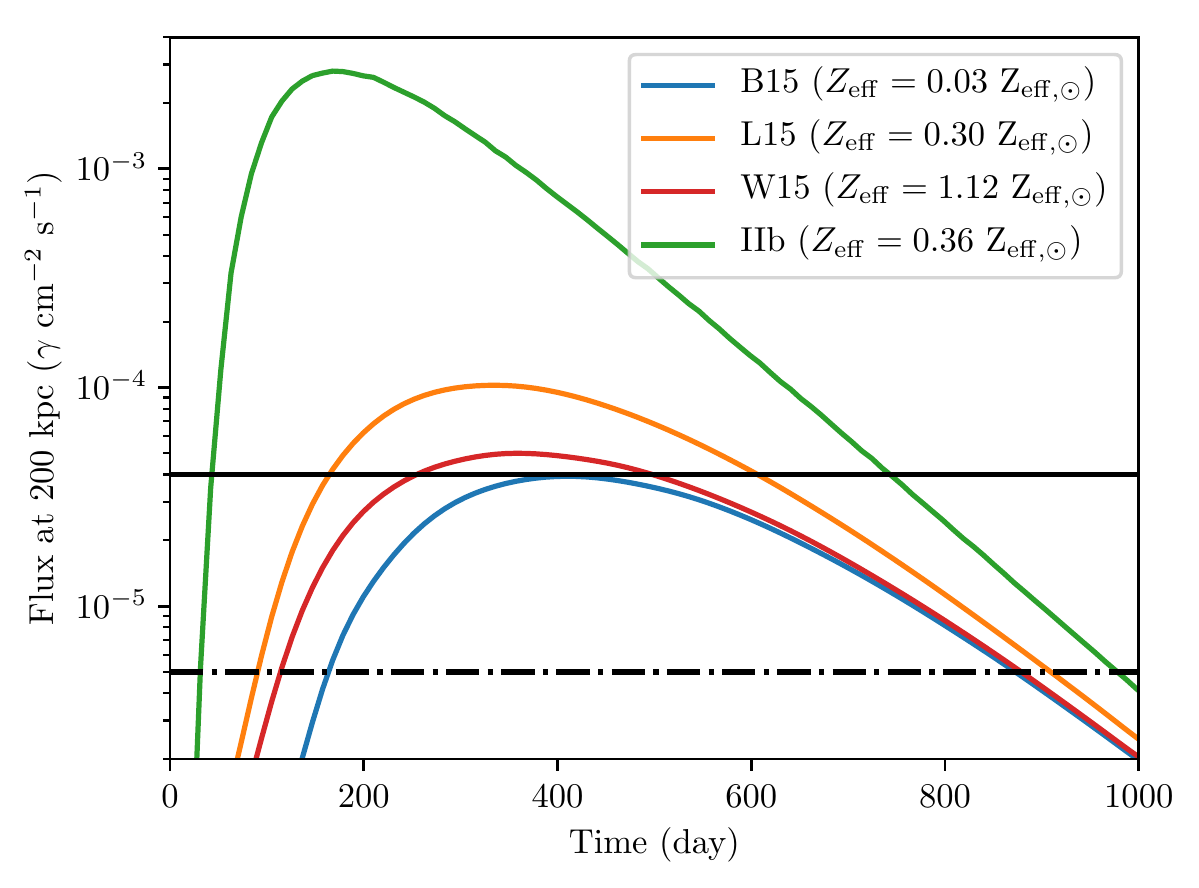}
  \caption{The 847~keV line light curves scaled to 200~kpc overplotted
    on the detection sensitivities of \textit{INTEGRAL}/SPI (solid
    black line, \citealt{roques03}) and the ESA M5 proposal
    \astrogam{} (dash-dotted black line, \citealt{de_angelis17}). The
    sensitivities are given for a detection threshold of 3$\sigma$ and
    an exposure time of 1~Ms.\label{fig:sensitivity_line}}
\end{figure}
Figure~\ref{fig:sensitivity_line} shows the computed 847~keV line
light curves and the narrow-line sensitivities of \textit{INTEGRAL}
and \astrogam{}. It is clear that \textit{INTEGRAL} is only capable of
detecting the 847~keV line out to around 200~kpc for the non-stripped
SNe. This effectively limits the range to within the Milky Way and its
satellites. Stripped-envelope SNe are expected to be detectable out to
2~Mpc, which covers the Local Group. \astrogam{} should expand the
horizon by a factor of three, which would increase the number of
potential targets by a factor of ${\sim}$30.

The expected CCSN rate is around 0.1 per year within 3~Mpc and 1 per
year within 10~Mpc \citep{arnaud04, ando05, botticella12, horiuchi13,
  xiao15}.  Inferred rates based on galaxy properties and star
formation models are associated with uncertainties. Optical surveys,
however, are possibly also incomplete by ${\sim}$20\,\% even within
10~Mpc \citep{prieto12, jencson17, jencson18, tartaglia18}. The
fraction of stripped-envelope SNe is estimated to be in the range
0.25--0.35, with reasonable agreement between estimates based on local
SNe observed over the past decades \citep{mattila12, botticella12,
  horiuchi13, xiao15} and surveys of larger volumes \citep{smartt09b,
  arcavi10, li11, smith11}. Thus, a reasonable estimate is that a CCSN
should be detectable by \textit{NuSTAR} every three years and the most
likely candidates are stripped-envelope SNe. Finally, we stress that
simply comparing predicted spectra with sensitivity curves only
provides a very rough estimate of what can be detected. Simulations
that include detailed instrumental effects and realistic backgrounds
will be the subject of future studies.

%%%%%%%%%%%%%%%%%%%%%%%%%%%%%%%%%%%%%%%%%%%%%%%%%%%%%%%%%%%%%%%%
\section{Summary \& Conclusions}\label{sec:conclusions}
We use SN models based on 3D neutrino-driven explosion simulations
\citep[][\gabler{}]{wongwathanarat13, wongwathanarat15,
  wongwathanarat17} to compute the expected early X-ray and gamma-ray
emission. Four of the models are designed to represent \sna{}; two of
these are single-star BSGs and two are the results of mergers. We
compare predictions from these models to observations of \sna{} to
constrain the model properties. Additionally, we investigate models of
two single-star RSGs and a stripped-envelope IIb model to extend the
results to other types of \mbox{CCSNe} that are more common than
\sna{}-like events. Our main conclusions are as follows:
\begin{enumerate}
\item The overall agreement between observations and model predictions
  indicates that the delayed neutrino-heating mechanism is able to
  produce SN explosions that are basically consistent with the X-ray
  and gamma-ray observations. General features are well reproduced,
  such as the normalization, spectral shape, and shape of the light
  curves.  We stress that these models are based on realistic
  simulations of the progenitors and SN explosions.
\item Both the single-star progenitor B15 and the merger model \ma{}
  are capable of reproducing the most relevant observational X-ray and
  gamma-ray properties of \sna{}. \ma{}, however, is the only
  progenitor out of the six merger models of \citet{menon17} that is
  able to match the main features of the observations. Similarly, the
  single-star model N20 can be excluded. The primary reason for
  failing to match the observations is insufficient mixing of \nc{} to
  the outer layers, which is related to the growth rates of
  Rayleigh-Taylor instabilities during the explosions
  \citep[][\utrobin{}]{wongwathanarat15, utrobin19}. This also
  highlights that X-ray and gamma-ray observations are a powerful way
  of constraining progenitor models.
\item On a more detailed level, there are differences in the temporal
  evolution of the continuum and line fluxes between the 3D explosion
  models and \sna{} observations. A suitable choice of viewing angle
  is not sufficient to reconcile these shortcomings. The differences
  can, however, be remedied by relatively small changes to the
  explosion dynamics. Thus, we do not consider these discrepancies to
  be critical issues in the explosion mechanism. Rather, they may
  potentially provide further insight to refine the progenitor
  models. For example, relative to the B15 model, it is sufficient to
  increase the velocity of the fastest trace amounts of \nc{} on the
  near side by ${\sim}$20\,\%, while the bulk of the \nc{} is
  redshifted by ${\sim}$500\kmps{} instead of ${\sim}$200\kmps{}. This
  is only one possible explanation, which illustrates that relatively
  small changes to the explosion dynamics and progenitor structures
  are needed, especially considering the sensitivity of the dynamics
  to the progenitor structure. This issue will be further discussed in
  a follow-up paper by \jerkstrandt{}.
\item The low-energy spectral cutoff is determined by the
  photoabsorption opacity of the progenitor envelope around 30~keV. In
  our explosion models, the outer parts of the envelopes are
  relatively spherical, which means that the low-energy X-ray cutoff
  is insensitive to viewing angle. This is potentially a direct way of
  observationally constraining the composition of SN progenitors. The
  observations of \sna{} are only weakly constraining and we find that
  the metallicity of its progenitor is consistent with both the
  metallicity of the LMC as well as the metallicity of its equatorial
  ring.
\item The asymmetries and 3D structures introduce a viewing-angle
  dependence, which primarily affects the overall flux
  normalization. For the more asymmetric models, the shapes of the
  light curves also change significantly for different viewing
  angles. The shapes of the spectra, however, remain relatively
  unaffected. The magnitude of these effects varies significantly
  depending on the level of ejecta asymmetries, epoch, and energy
  range considered.
\item The most important properties that affect the nature of the
  X-ray and gamma-ray emission are the amount of \nc{} mixing and the
  level of asymmetry. Aside from this, qualitatively similar
  progenitor models produce relatively similar X-ray and gamma-ray
  emission. The X-ray and gamma-ray emission of the stripped-envelope
  IIb model evolves faster and is more than an order of magnitude more
  luminous than the non-stripped models.
\item \textit{NuSTAR} offers the best prospects of future observations
  of early X-ray continuum emission from nearby SNe. Based on simple
  estimates, it should be capable of detecting non-stripped SNe within
  3~Mpc and stripped-envelope SNe out to 10~Mpc, which extends to the
  nearest galaxies beyond the Local Group. This corresponds to an
  expected detection rate of 1 CCSN every three years. The deepest
  observations of direct line emission among the current instruments
  are provided by \textit{INTEGRAL}/SPI. It is expected to cover
  non-stripped SNe in the Milky Way and its satellites, and reach
  stripped SNe at 2~Mpc, which is comparable to the extent of the
  Local Group.
\end{enumerate}

\acknowledgments{We thank Markus Kromer, Lih-Sin The, and Peter Milne
  for providing the W7 model \citep{nomoto84b}, and Stan Woosley for
  providing the 10HMM model \citep{pinto88b}. This research was funded
  by the Knut \& Alice Wallenberg foundation. K.M. acknowledges
  support by JSPS KAKENHI Grant (18H04585, 18H05223, 17H02864). The
  simulations were in part carried out on Cray XC30/XC50 at the Center
  for Computational Astrophysics, National Astronomical Observatory of
  Japan. At Garching, this project was supported by the European
  Research Council through grant ERC-AdG No. 341157-COCO2CASA, and by
  the Deutsche Forschungsgemeinschaft through Sonderforschungbereich
  SFB 1258 ``Neutrinos and Dark Matter in Astro- and Particle
  Physics'' (NDM) and the Excellence Cluster Universe (EXC 153;
  \url{http://www.universe-cluster.de/}). The 3D explosion modeling
  was performed using the draco and cobra clusters at the Max Planck
  Computing and Data Facility (MPCDF). A.H. has been supported, in
  part, by a grant from Science and Technology Commission of Shanghai
  Municipality (Grants No.16DZ2260200) and National Natural Science
  Foundation of China (Grants No.11655002), and by the Australian
  Research Council through a Future Fellowship (FT120100363). This
  research has made use of NASA's Astrophysics Data System.}

\vspace{5mm}
\facilities{\textit{SMM} (GRS), \textit{Ginga} (LAC)}
\software{
  \texttt{astropy} \citep{astropy13},
  \texttt{FTOOLS} \citep{blackburn95},
  \texttt{h5py} \citep{collette13},
  \texttt{HEAsoft} \citep{heasarc14},
  \texttt{matplotlib} \citep{hunter07},
  \texttt{numpy} \citep{jones01, van_der_walt11},
  \textsc{Prometheus} \citep{fryxell91, muller91},
  \textsc{Prometheus-HOTB} (\citealt{wongwathanarat10b}; \gabler{}),
  \textsc{Prometheus-Vertex} \citep{rampp02, buras06},
  \texttt{SAOImage DS9} \citep{joye03},
  \texttt{scipy} \citep{jones01}.
}

\bibliography{references}
\end{document}